\documentclass[12pt]{article}

\usepackage{amsmath}
\usepackage{amsfonts}
\usepackage{amssymb}
\usepackage{graphicx}
\input amssym.def
\input amssym
\newcommand{\be}{\begin{equation}}
\newcommand{\ee}{\end{equation}}

\begin{document}

\renewcommand{\theequation}{\arabic{section}.\arabic{equation}}

\title{BF systems on graph cobordisms as topological cosmology}
\author{Vladimir N. Efremov\thanks{Mathematics Department, CUCEI,
University of Guadalajara, Guadalajara, Mexico.}
\thanks{E-mail:
efremov@udgserv.cencar.udg.mx}\\
Nikolai V. Mitskievich\thanks{Physics Department, CUCEI,
University of Guadalajara, Guadalajara, Mexico.}
\thanks{E-mail:
mitskievich03@yahoo.com.mx}\\
Alfonso M. Hern\'andez Magdaleno$^\ddag$ }
\date{~}

\maketitle


\begin{abstract}
A cosmological model connecting the evolution of universe with a
sequence of topology changes described by a collection of specific
graph cobordisms, is constructed. It is shown that an adequate
topological field theory (of BF-type) can be put into relation to
each graph cobordism. The explicit expressions for transition
amplitudes (partition functions) are written in these BF-models
and it is shown that the basic topological invariants of the graph
cobordisms (intersection matrices) play the r{\^o}le of coupling
constants between the formal analogues of electric and magnetic
fluxes quantized {\`a} la Dirac, but with the use of
Poicar{\'e}--Lefschetz duality. For a specific graph cobordism,
the diagonal elements and eigenvalues of the intersection matrix
reproduce the hierarchy of dimensionless low-energy coupling
constants of the fundamental interactions acting in the real
universe.\\

\end{abstract}

\newpage
\section{Introduction} \label{s1} \setcounter{equation}{0}
The main hypothesis we advance in this paper is that the spacetime
to\-po\-lo\-gy determines ({\it via} an Abelian topological
BF-type field theory) the number and hierarchy of coupling
constants of the fundamental (pre-)interactions which are adequate
to the topological structure of the real universe. Thus we
continue to develop the principal idea of \cite{EM} that the
primary values of coupling constants (which correspond to a vacuum
without any local excitations) are topological invariants of
four-dimensional spacetime manifold. The topological field
theories \cite{Witt88, BlauThom, Hor, BaezMG} are in fact
exercises in calculation of topological invariants (the ``theory
of nothing'') \cite{Thom} accompanied by certain physical
interpretations. In particular, in our $U(1)$ gauge BF-model (in
the vein of works on global aspects of electric-magnetic duality
\cite{Verl, WitS, OlAl1, OlAl2, OlAl3, Zu1, Zu2}) a determination
of the complete system of the gauge classes of BF system solutions
(phase space) is equivalent to specification of topology of
spacetime $M$ because of the existence of isomorphisms of type
$$
\textnormal{Princ}\,(M)\overset{c}{\cong}H^2(M,\mathbb{Z})
$$
between the group of principal $U(1)$-bundles over $M$ and the
group of cohomology classes of 2-cycles of $M$ (see the formulae
(\ref{2.10}) -- (\ref{2.13})) and since the groups of
cohomological classes of cocycles with dimensions 1 and 3 are
trivial for the set of four-dimensional manifolds under
consideration, namely graph cobordisms, which are intensively
studied by mathematicians in recent years \cite{FFU, SavArt,
NemNicol, NeumWahl}.

The problem is set as in self-consistent cosmological model
building: to construct a spacetime manifold  which admits fields
configurations with non-zero fluxes through the collection of
homologically non-trivial 2-cycles \cite{Verl}, {\it i.e.} closed
surfaces that do not correspond to the boundary of a
three-dimensional submanifold in $M$. A generalization of Dirac's
quantization conditions \cite{Verl, WitS, OlAl2, OlAl3, Zu2}
(formulated on the basis of the Poincar{\'e}--Lefschetz duality)
implies that these fluxes must be quantized. We show that the
coupling constant matrix describing interactions between the
quantized fluxes, coincides up to a scale factor (which also is
found to be a topological invariant) with the basic topological
invariant of spacetime $M$, {\it i.e.} with its intersection
matrix $Q$. For the specially constructed graph manifold
$M^+_D(0)$ (see section 4), diagonal elements of the intersection
matrix $Q^+(0)$ reproduce rather exactly the hierarchy of
dimensionless low-energy coupling (DLEC) constants of the
universal physical interactions acting in our universe. The other
graph cobordisms $M^+_D(t)$, $t=-1,-2,-3,-4$, constructed in the
same section, probably describe earlier stages of cosmological
evolution, which are distinguished from each other rather by the
spacetime topology than by gauge groups of physical fields (the
gauge group we use is $U(1)$). Thus our model demonstrates that
topological invariants may codify information about the physical
interactions, which can be naturally introduced on a non-trivial
topological space forming the spacetime background of the
self-consistent cosmological model which involves qualitatively
different evolutionary phases connected one to another by topology
changes.

The relations between the topology fluctuations and the problem of
fixing fundamental coupling constants are discussed widely, see
for example \cite{Pres89, Wein89}. In particular, the solution of
the cosmological-constant problem in terms of spontaneous topology
changes is treated in \cite{Haw88, Col88, GidStr, KlSuBa}. We
propose a fairly different approach to these old problems by means
of using cobordisms of sufficiently specific type to model the
spacetime manifolds.

We remind that if the boundary of a compact four-dimensional
topological space $M$, $\partial M$, is a disjoint sum of two
three-dimensional topological spaces $\Sigma_{\rm{in}}$ and
$\Sigma_{\rm{out}}$, the 3-tuple $(M,\Sigma_{\rm{in}},
\Sigma_{\rm{out}})$ is called a cobordism \cite{RourSan}. Both
$\Sigma_{\rm{in}}$ and $\Sigma_{\rm{out}}$ are often merely
implied, thus the very topological space $M$ is called a
cobordism.

The paper is organized as follows. The section \ref{s2} contains
the basic mathematical concepts which are used to construct a
topological field theory on graph cobordisms. In the subsection
\ref{s2.1} we reproduce the definitions of spicing and plumbing
operations  which are necessary to glue graph cobordisms according
to algorithms codified by means of splice diagrams. These concepts
are unussual for physicists, but are well developed by topologists
\cite{Hirz, EisNeum, SavB2}. The subsection \ref{s2.2} contains a
review of the basic notions connected with the principal
$U(1)$-bundles on four manifolds with nonempty boundaries in the
style of \cite{Zu2}. Here we also discuss the non-trivial (but
rather simple) (co-) homological properties of graph cobordisms.
The following subsection (\ref{s2.3}) is dedicated to a brief
survey of Poincar{\'e}--Lefschetz duality \cite{SavB1, SavB2}
which we use instead of the common Hodge duality to define the
formal analogues of electric and magnetic fluxes in our version of
BF-theory. Moreover we  remind  the definition of the main
topological invariant of four-dimensional manifolds, namely
intersection form \cite{Hirz, EisNeum} which serves for the
determination of coupling constant matrices characterizing
interactions of these ``electric'' and ``magnetic'' fluxes.

In the section \ref{s3} we construct the simple version of the
Abelian BF-theory on graph cobordisms, also known as plumbed
$V$-cobordisms \cite{SavArt}. In analogy with the electrodynamics
with theta term \cite{Verl, WitS, Zu2}, the ``electric'' and
``magnetic'' fluxes are defined as linear combinations of the
first Chern classes of graph cobordism $M$ and its boundary
$\partial M$. Then the transition amplitudes (partition functions)
are  expressed as functionals of these fluxes, intersection
matrices and the BF scale factor $\lambda$. These transition
amplitudes are topological invariants and represent something
resembling to the theta function. They also support a certain form
of strong-weak coupling duality.

In the section \ref{s4} we present numerical calculations of
intersection matrices for a specific sequence of graph cobordisms
which can be interpreted as a series of  the topology changes
leadind  to a certain state of the universe. This state can be
identified with the contemporary one by means of the elementary
interactions between ``electric'' fluxes, since the coupling
constants hierarchy of these fluxes reproduces the hierarchy of
the fundamental interactions in the real universe. At the end of
this section we give interpretation and discussion of the obtained
results.

The standard notations ${\mathbb Z}$, ${\mathbb R}$ and ${\mathbb
C}$ are used for the sets of integer, real and complex numbers,
respectively.

\section{Mathematical concepts} \label{s2} \setcounter{equation}{0}

As this was said in the Introduction, we build here a rather
simple topological gauge (BF-) model on sufficiently complicated
topological spaces belonging to the class of graph cobordisms
imitating tunnelling topological changes.
These four-dimensional smooth manifolds with the Euclidean
signature possess non-trivial (co-)homological characteristics.
This leads to a specific generalization of the Dirac quantization
conditions \cite{Verl, WitS, OlAl2} and enables us to explicitly
express transition amplitudes (partition functions) in terms of
the topological invariants (intersection forms) of the
graph cobordisms. The boundary components of
graph cobordisms are disjoint sums of lens spaces and
$\mathbb{Z}$-homology spheres. First we give some necessary
definitions following the works of Saveliev \cite{SavB2, SavArt},
see also \cite{EisNeum}.

\subsection{$\mathbb{Z}$-homology spheres, splicing, plumbing and
graph cobordisms} \label{s2.1}

Let $a_1,~a_2,~a_3$ be pairwise relatively prime positive numbers.
The Brieskorn homology sphere (Bh-sphere) $\Sigma(\underline{a})
:=\Sigma(a_1,a_2,a_3)$ is defined as the link of (Brieskorn)
singularity \be \label{2.1} \Sigma(\underline{a}):=\Sigma(a_1,
a_2,a_3):=\left\{{z_1}^{a_1}+{z_2}^{a_2} +{z_3}^{a_3}=0\right\}
\cap S^5 \ee where $z_i\in\mathbb{C}_i$, and $S^5$ is the unit
five-dimensional sphere $|z_1|^2+|z_2|^2+|z_3|^2=1$. The singular
complex algebraic surface ${z_1}^{a_1}+{z_2}^{a_2} +{z_3}^{a_3}=0$
has the canonical orientation which induces the canonical
orientation of the link $\Sigma(\underline{a})$. If any of $a_i$
is equal to 1, the manifold $\Sigma(\underline{a})$ is
homeomorphic to the ordinary $S^3$. Bh-spheres belong to the class
of Seifert fibered homology (Sfh-) spheres \cite{Neum77}. On this
manifold, there exists a unique Seifert fibration which has
unnormalized Seifert invariants \cite{NeumRaym} $(a_i,b_i)$
subject to $e(\Sigma (\underline{a}))= \sum_{i=1}^{3} b_i/a_i
=1/a$, where $a=a_1 a_2 a_3$ and $e(\Sigma (\underline{a}))$ is
its Euler number (the well known topological invariant of a
Bh-sphere). This Seifert fibration is defined by the $S^1$-action
which reads $t(z_1, z_2, z_3)= (t^{\sigma_1}z_1, t^{\sigma_2}z_2,
t^{\sigma_3}z_3)$, where $t\in S^1$, and $\sigma_i= a/a_i$. This
action is fixed-point-free. The only points of $\Sigma(a_1,
a_2,a_3)$ which have non-trivial isotropy group $\mathbb{Z}_{a_i}$
are those with one coordinate $z_i$ equal to 0 ($i=1,2,3$). The
fiber through such a point is called an exceptional (singular)
fiber of degree $a_i$. All other fibers are called regular
(non-singular). In general, Sfh-spheres $\Sigma(a_1,...,a_n)$ have
$n$ different exceptional fibers and represent special cases of
$\mathbb{Z}$-homology spheres \cite{EisNeum}.

By a $\mathbb{Z}$-homology sphere we mean a closed three-manifold
$\Sigma$ such that all homology groups of $\Sigma$ with integer
coefficients are isomorphic to homology groups of the ordinary
three-sphere $S^3$ over $\mathbb{Z}$. All $\mathbb{Z}$-homology
spheres used in this paper can be obtained from Bh-spheres by the
splicing operation. This operation is defined for any Sfh-sphere
as follows: First we define \cite{EisNeum} a Seifert link as a
pair $(\Sigma,S)=(\Sigma,S_1\bigcup\cdots \bigcup S_m)$ consisting
of oriented $\mathbb{Z}$-homology sphere $\Sigma$ and a collection
$S$ of Seifert fibers (exceptional or regular) $S_1,\dots,S_m$ in
$\Sigma$. Note that the links $(S^3,S)$ where $S^3$ is an ordinary
three-sphere, are also allowed. Let $(\Sigma,S)$ and
$(\Sigma',S')$ be links and choose components $S_i\in S$ and
$S'_j\in S'$. Let also $N(S_i)$ and $N(S'_j)$ be their tubular
neighbourhoods, while $m,l\subset\partial N(S_i)$ and
$m',l'\subset\partial N(S'_j)$ be standard meridians and
longitudes. The manifold
$\Sigma''=(\Sigma\setminus\textnormal{int}N(S_i))
\bigcup(\Sigma'\setminus\textnormal{int}N(S'_j))$ obtained by
pasting along the torus boundaries by matching $m$ to $l'$ and
$m'$ to $l$, is a ${\Bbb Z}$-homology sphere. The link $\left(
\Sigma'',(S\setminus S_i)\bigcup(S'\setminus S'_j)\right)$ is
called the splice (splicing) of $(\Sigma,S)$ and $(\Sigma',S')$
along $S_i$ and $S'_j$. We shall use the standard notation
$\Sigma''= \Sigma\frac{~}{S_i ~ S'_j}\Sigma'$ or simply $\Sigma''=
\Sigma\frac{~ ~}{~ ~ ~}\Sigma'$. Any link which can be obtained
from a finite number of Seifert links by splicing is called a
graph link. Empty graph links are precisely the (graph) $\mathbb{
Z}$-homology spheres. All graph links are classified as in
\cite{EisNeum} by their splice diagrams.

A splice diagram $\Delta$ is a finite tree graph with vertices of
three types. Vertices with at least three adjacent edges are
called {\it nodes}. Each node with $n$ adjacent edges corresponds
to a Sfh-sphere $\Sigma(a_1,...,a_n)$. The edges adjacent to a
node correspond to exceptional fibers and are weighted by
$a_1,...,a_n$, respectively. The very node carries a sign plus if
$\Sigma(a_1,...,a_n)$ is oriented as a link of singularity of the
type (\ref{2.1}) and minus otherwise. Two other types of vertices
have only one adjacent edge and are called either {\it leaves} or
{\it arrowheads}. The former ones represent singular fibers in a
Sfh-sphere, while the latter ones, components of a graph link. We
shall use splice diagrams $\Delta_r$ with nodes of valence $n=3$
only, see figure 1 . This type of splice diagrams corresponds to
splicing of $r$ Bh-spheres. We just consider the splice diagrams
with pairwise coprime positive weights around each node. In this
case the concepts of integer ($\mathbb{Z}$-) homology spheres and
graph homology spheres are equivalent.

\begin{figure}[h]
\begin{center}

\setlength{\unitlength}{1pt}
\begin{picture}(150,50)
\put(-58,0){\line(1,0){35}}
\put(-18,5){\line(0,1){35}}
\put(-13,0){\line(1,0){45}}
\put(42,0){\line(1,0){45}}
\put(37,5){\line(0,1){35}}
\put(167,0){\line(1,0){35}}
\put(162,5){\line(0,1){35}}
\put(122,0){\line(1,0){35}}
\put(-18,38){\makebox(.5,.5){$\bullet$}}
\put(37,38){\makebox(.5,.5){$\bullet$}}
\put(162,38){\makebox(.5,.5){$\bullet$}}
\put(-58,-0.5){\makebox(.5,.5){$\bullet$}}
\put(202,-0.5){\makebox(.5,.5){$\bullet$}}
\multiput(93,-0.5)(10,0){3}{.}
\put(-18,0){\makebox(.5,.5){$+$}}
\put(-18,0){\circle{10}}
\put(37,0){\makebox(.5,.5){$+$}}
\put(37,0){\circle{10}}
\put(162,0){\makebox(.5,.5){$+$}}
\put(162,0){\circle{10}}
\put(-8,15){\makebox(.5,.5){$a_{11}$}}
\put(47,15){\makebox(.5,.5){$a_{12}$}}
\put(172,15){\makebox(.5,.5){$a_{1r}$}}
\put(-33,-7){\makebox(.5,.5){$a_{21}$}}
\put(-3,-7){\makebox(.5,.5){$a_{31}$}}
\put(22,-7){\makebox(.5,.5){$a_{32}$}}
\put(52,-7){\makebox(.5,.5){$a_{22}$}}
\put(147,-7){\makebox(.5,.5){$a_{3r}$}}
\put(177,-7){\makebox(.5,.5){$a_{2r}$}}
\end{picture}\\
\vspace*{1.cm}
\caption{A splice diagram $\Delta_r$.}
\end{center}
\end{figure}
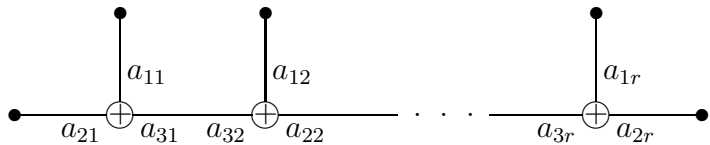

In general, graph homology spheres can be conveniently described
by plumbing. Plumbing graphs are required for introduction of
four-dimensional manifolds with graph homology spheres as
boundaries. The plumbing representation make it also possible to
define intersection forms of these four-manifolds and to pass to
the definition of graph cobordisms.

A {\it plumbing graph} $\Gamma$ is a graph with no cycles (a
finite tree) each of whose vertices $v_i$ carries an integer
weight $e_i, ~ i=1,\dots,r$. To each vertex $v_i$ a $D^2$-bundle
$Y(e_i)$ over $S^2$ is associated, whose Euler class
(self-intersection number of zero-section) is $e_i$. If the vertex
$v_i$ has $d_i$ edges connected to it on the graph $\Gamma$,
choose $d_i$ disjoint discs in the base $S^2$ of $Y(e_i)$ and call
the disc bundle over the $j$th disc $B_{ij}=\left(D^2_j\times
D^2\right)_i$. When two vertices $v_i$ and $v_k$ are connected by
an edge, the disc bundles $B_{ij}$ and $B_{kl}$ should be
identified by exchanging the base and fiber coordinates
\cite{Orlik}. This pasting operation is called {\it plumbing}, and
the resulting smooth four-manifold $P(\Gamma)$ is known as {\it
graph manifold} ({\it plumbed four-manifold}). Its boundary
$\Sigma(\Gamma)=\partial P(\Gamma)$ is referred to as a {\it
plumbed three-manifold}.

Since the homology group $H_1(P(\Gamma),\mathbb{Z})=0$, the unique
non-trivial homology characteristic is $H_2(P(\Gamma),\mathbb{Z})$
which has a natural basis (set of generators) represented by the
zero-sections of the plumbed bundles. All these sections are
embedded 2-spheres $z_i\mbox{ where }i=1,\dots,r=\mbox{ rank
}H_2(P(\Gamma),\mathbb{Z})$, and they can be oriented in such a
way that the intersection (bilinear) form \cite{SavB2} \be Q:
~H_2(P(\Gamma),\mathbb{Z})\otimes H_2(P(\Gamma),\mathbb{Z})
\rightarrow \mathbb{Z} \label{2.2} \ee will be represented by the
$r\times r$-matrix $Q(\Gamma)= (q_{ij})$ with the entries:
$q_{ij}=e_i$ if $i=j$; $q_{ij}=1$ if the vertex $v_i$ is connected
to $v_j$ by an edge; and $q_{ij}=0$ otherwise. The three-manifold
$\Sigma(\Gamma)$ is $\mathbb{Z}$-homology sphere iff the matrix
$Q(\Gamma)$ is unimodular, that is det\,$Q(\Gamma)=1$.

In order to construct a plumbing representation for a
$\mathbb{Z}$-homology sphere given by a splice diagram $\Delta$,
we need two things:
\begin{enumerate}
\item Plumbing graphs for the basic building blocks, {\it i.e.}
Sfh-spheres (in our case, Bh-spheres). \item A procedure to splice
together plumbing graphs.
\end{enumerate}
First, let $\Sigma$ be a Sfh-sphere with unnormalized Seifert
invariants $(a_i,b_i)$, $i=1,...,n$, and the splice diagram of
figure 2. It can be obtained as a boundary of the graph manifold
$P(\Gamma)$ where $\Gamma$ is a star graph shown in figure 3
\cite{Orlik}. The integer weights $t_{ij}$ in this graph are found
from continued fractions $a_i/b_i=[t_{i1},\dots,t_{im_i}]$; here
$$
[t_1,\dots,t_k]=t_1-\frac{1}{t_2-\displaystyle\frac{1}{\cdots-
\displaystyle\frac{1}{t_k}}}.
$$
\begin{figure}[h]
\begin{center}
\setlength{\unitlength}{1pt}
\begin{picture}(150,50)

\put(63,14){\line(1,1){20}}
\put(63,6){\line(1,-1){20}}

\put(84,34.5){\makebox(.5,.5){$\bullet$}}
\put(84,-15.5){\makebox(.5,.5){$\bullet$}}

\multiput(75,1)(0,9){3}{.}

\put(60,10){\makebox(.5,.5){$+$}}
\put(60,10){\circle{10}}

\put(65,30){\makebox(.5,.5){$a_{1}$}}
\put(65,-10){\makebox(.5,.5){$a_{n}$}}

\end{picture}\\
\vspace*{1.cm} \caption{The splice diagram
of an Sfh-sphere $\Sigma$.}
\end{center}
\end{figure}
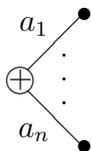

Lens spaces represent a special case of Seifert fibered manifolds.
Expanding $-p/q=[t_1,\dots,t_n]$ into a continued fraction, we
encounter $L(p,q)$ as a boundary of the 4-manifold obtained by
plumbing on the chain $\Gamma^{\rm ch}_i$ shown in figure 4.

\begin{figure}[h]     
\begin{center}
\setlength{\unitlength}{1pt}
\begin{picture}(150,80)
\put(29,60){\line(1,0){65}} \put(29,00){\line(1,0){65}}
\put(115,60){\line(1,0){25}} \put(115,00){\line(1,0){25}}
\put(0,30){\line(1,1){30}} \put(0,30){\line(1,-1){30}}
\put(30,59.5){\makebox(.5,.5){$\bullet$}}
\put(30,-.5){\makebox(.5,.5){$\bullet$}}
\put(95,57){\makebox(20,5){...}} \put(95,-3){\makebox(20,5){...}}
\put(140,59.5){\makebox(.5,.5){$\bullet$}}
\put(140,-.5){\makebox(.5,.5){$\bullet$}}
\put(70,59.5){\makebox(.5,.5){$\bullet$}}
\put(70,-.5){\makebox(.5,.5){$\bullet$}}
\put(0,29.5){\makebox(.5,.5){$\bullet$}}
\put(-9,29.5){\makebox(.5,.5){$0$}}
\put(30,69.5){\makebox(.5,.5){$t_{11}$}}
\put(30,-10.5){\makebox(.5,.5){$t_{n1}$}}
\put(140,69.5){\makebox(.5,.5){$t_{1m_1}$}}
\put(140,-10.5){\makebox(.5,.5){$t_{nm_n}$}}
\put(70,69.5){\makebox(.5,.5){$t_{12}$}}
\put(70,-10.5){\makebox(.5,.5){$t_{n2}$}}
\put(55,12){\makebox(20,5){...}} \put(95,12){\makebox(20,5){...}}
\put(55,27){\makebox(20,5){...}} \put(95,27){\makebox(20,5){...}}
\put(55,42){\makebox(20,5){...}} \put(95,42){\makebox(20,5){...}}
\end{picture}\\
\vspace*{1.cm}
\caption{The star graph $\Gamma$.}
\end{center}
\end{figure}
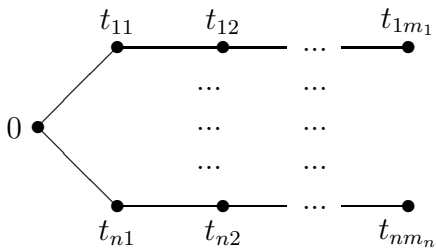

Notice that this plumbing graph simultaneously represents the lens
space $L(p,q^*)$ with $-p/q^*=[t_n,\dots,t_1]$ where $qq^*=1$ mod
$p$. This reflects the fact that $L(p,q)$ and $L(p,q^*)$ are
homeomorphic. Moreover, to the continuous fraction
$a_i/b_i=[t_{i1},\dots,t_{im_i}]$ there correspond both the
subgraph $\Gamma^{\rm ch}_i$ of $\Gamma$ shown in figure 4, and
the lens space $L(-a_i,b_i)$ (also called {\it leaf lens space}).

\begin{figure}[h]     
\begin{center}
\setlength{\unitlength}{1pt}
\begin{picture}(150,15)
\put(2,00){\line(1,0){89}} \put(117,00){\line(1,0){23}}
\put(3,-.5){\makebox(.5,.5){$\bullet$}}
\put(95,-3){\makebox(20,5){...}}
\put(140,-.5){\makebox(.5,.5){$\bullet$}}
\put(50,-.5){\makebox(.5,.5){$\bullet$}}
\put(4.5,-10.5){\makebox(.5,.5){$t_{1}$}}
\put(141.5,-10.5){\makebox(.5,.5){$t_{n}$}}
\put(51.5,-10.5){\makebox(.5,.5){$t_{2}$}}
\end{picture}\\
\vspace*{1.cm} \caption{One-dimensional
chain $\Gamma^{\rm ch}_i$.}
\end{center}
\end{figure}
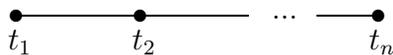

Next, we address the problem of splicing together plumbing graphs.
It can be described as follows. Suppose that two graph links are
represented by their plumbing diagrams $\overline{\Gamma}$ and
$\overline{\Gamma'}$ (see figure 5) with arrows attached to
vertices $e_n$ and $e'_m$, respectively. The corresponding
plumbing diagram for a spliced link is shown in figure 6  where
$a=\det Q(\Gamma_0)/\det Q(\Gamma)$, while $\Gamma$ is the
plumbing graph $\overline{\Gamma}$ with the arrow deleted, and
$\Gamma_0$ is a portion of $\Gamma$ obtained by removing the $n$th
vertex weighted by $e_n$ as well as all its adjacent edges.
Another integer $a'$ is similarly obtained from the graph
$\Gamma'$ (examples see in \cite{EisNeum,SavArt}). The above
description of splicing in terms of plumbing graphs makes it
possible to treat splicing as an operation on the corresponding
plumbed 4-manifold; moreover, $\Sigma(\Gamma)=\partial P(\Gamma)$.

\begin{figure}[h]     
\begin{center}
\setlength{\unitlength}{1pt}
\begin{picture}(150,50)
\put(2,00){\line(1,0){45}} \put(140,00){\line(-1,0){45}}
\multiput(2,-0.5)(-3,-3){10}{.}   
\multiput(2,-0.5)(-3,3){10}{.}     
\multiput(140,0)(3,3){10}{.} \multiput(140,0)(3,-3){10}{.}
\put(47,0){\vector(1,0){5}} \put(95,0){\vector(-1,0){5}}
\put(5,-15){\makebox(10,5){$e_n$}}
\put(135,-15){\makebox(10,5){$e'_m$}}
\put(140,-.5){\makebox(.5,.5){$\bullet$}}
\put(2,-.5){\makebox(.5,.5){$\bullet$}}
\end{picture}\\
\vspace*{1.cm} \caption{Plumbing diagrams $\overline{\Gamma}$ and
$\overline{\Gamma'}$ prepared for splicing.}
\end{center}
\end{figure}
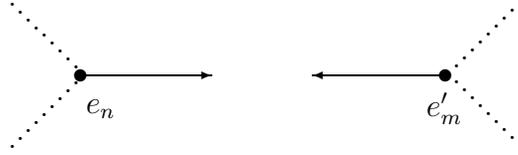

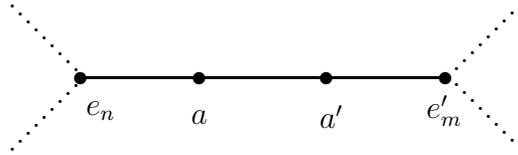
\begin{figure}[h]     
\begin{center}
\setlength{\unitlength}{1pt}
\begin{picture}(150,50)
\put(2,00){\line(1,0){140}} 
\multiput(2,-0.5)(-3,-3){10}{.}   
\multiput(2,-0.5)(-3,3){10}{.}     
\multiput(140,0)(3,3){10}{.} \multiput(140,0)(3,-3){10}{.}
\put(47,-0.5){\makebox(.5,.5){$\bullet$}}
\put(95,-0.5){\makebox(.5,.5){$\bullet$}}
\put(5,-15){\makebox(10,5){$e_n$}}
\put(135,-15){\makebox(10,5){$e'_m$}}
\put(140,-.5){\makebox(.5,.5){$\bullet$}}
\put(2,-.5){\makebox(.5,.5){$\bullet$}}
\put(47,-15){\makebox(.5,.5){$a$}}
\put(97,-15){\makebox(.5,.5){$a'$}}
\end{picture}\\
\vspace*{1.cm}\caption{Plumbing diagram
obtained by splicing $\overline{\Gamma}$ and
$\overline{\Gamma'}$.}
\end{center}
\end{figure}

The important problem in calculation with plumbed graphs is to
identify the {\it extra lens space} arising between nodes in the
course of splicing. See figure 7 where the lens space in question
is decorated by oval. The resulting lens space $L(p,q)$
($-p/q=[t_1, \dots,t_k]$) is characterized by the following
parameters, \be \label{2.3} p=a_1\cdots
a_{n-1}\alpha_1\cdots\alpha_{m-1}-a_n \alpha_m, \ee \be
\label{2.4} q=-a_1\cdots
a_{n-1}\alpha_1\cdots\alpha_{m-1}\sum_{i=1}^{n-1}
\frac{b_i}{a_i}-b_n \alpha_m. \ee This lens space depends only on
the spliced Seifert links $\left(\Sigma(a_1,...,a_n),S_n \right)$
and $\left(\Sigma(\alpha_1,...,\alpha_m),S_m \right)$, but does
not depend on the rest of the splice diagram \cite{SavArt,
EisNeum}.

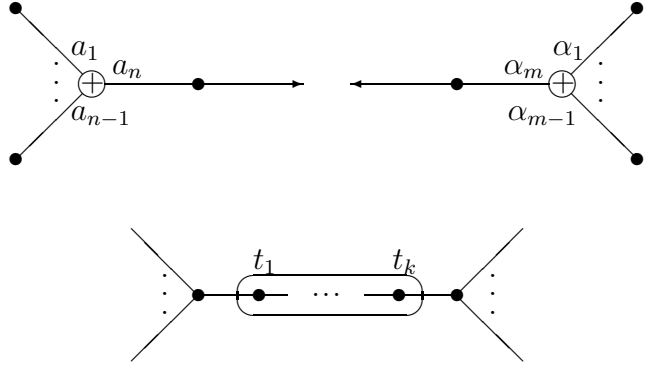
\begin{figure}[h]
\begin{center}
\setlength{\unitlength}{1pt}
\begin{picture}(150,50)

\put(22,00){\line(1,0){35}} \put(120,00){\line(-1,0){35}}
\put(57,0){\vector(1,0){5}} \put(85,0){\vector(-1,0){5}}
\put(-13,0){\line(1,0){35}}
\put(120,0){\line(1,0){35}}
\put(22,-.5){\makebox(.5,.5){$\bullet$}}
\put(120,-.5){\makebox(.5,.5){$\bullet$}}
\put(-18,0){\makebox(.5,.5){$+$}}
\put(-18,0){\circle{10}}
\put(160,0){\makebox(.5,.5){$+$}}
\put(160,0){\circle{10}}
\put(163.5,3.5){\line(1,1){25}}
\put(163.5,-3.5){\line(1,-1){25}}
\put(-21.5,-3.5){\line(-1,-1){25}}
\put(-21.5,3.5){\line(-1,1){25}}
\put(-47,28){\makebox(.5,.5){$\bullet$}}
\put(-47,-29){\makebox(.5,.5){$\bullet$}}
\put(188,28){\makebox(.5,.5){$\bullet$}}
\put(188,-29){\makebox(.5,.5){$\bullet$}}
\multiput(-33,-7)(0,7){3}{.}
\multiput(173,-7)(0,7){3}{.}
\put(-23,10){\makebox(5,5){$a_{1}$}}
\put(-7,3){\makebox(5,5){$a_{n}$}}
\put(-17,-15){\makebox(5,5){$a_{n-1}$}}
\put(160,10){\makebox(5,5){$\alpha_{1}$}}
\put(143,3){\makebox(5,5){$\alpha_{m}$}}
\put(150,-15){\makebox(5,5){$\alpha_{m-1}$}}

\put(21,-80){\line(1,0){35}} \put(120,-80){\line(-1,0){35}}
\put(45,-80.5){\makebox(.5,.5){$\bullet$}}
\put(98,-80.5){\makebox(.5,.5){$\bullet$}}
\put(22,-80.5){\makebox(.5,.5){$\bullet$}}
\put(120,-80.5){\makebox(.5,.5){$\bullet$}}
\multiput(65,-80.5)(4,0){3}{.}
\put(72,-80){\oval(70,15)}
\put(120,-80){\line(1,1){25}}
\put(120,-80){\line(1,-1){25}}
\put(22,-80){\line(-1,-1){25}}
\put(22,-80){\line(-1,1){25}}
\multiput(8,-87)(0,7){3}{.}
\multiput(132,-87)(0,7){3}{.}
\put(45,-70){\makebox(5,5){$t_{1}$}}
\put(98,-70){\makebox(5,5){$t_{k}$}}

\end{picture}\\
\vspace*{4.cm} \caption{The extra lens
space arising as a result of splicing.}
\end{center}
\end{figure}

If all extra lens spaces (which we also call ``lens spaces between
nodes'') $L(p'_k,q'_k)$, $k=1,...,N_{\rm extra}$, are subject to
the condition $p'_k<0$, then $\partial P(\Gamma)=\Sigma_{\rm alg}$
is an {\it algebraic link} \cite{EisNeum} which consequently
bounds the graph manifold $P(\Gamma)$ having a negative defined
intersection form $Q(\Gamma)$.

Now we are ready to define topological spaces of most importance
in this paper, the graph cobordisms also known as plumbed
V-cobordisms \cite{SavArt} related to the decorated plumbed
graphs. Let $P(\Gamma)$ be a plumbed four-manifold corresponding
to graph $\Gamma$, and $\Gamma^{\rm ch}$ be a chain in $\Gamma$ of
the form shown in Figure 4. Plumbing on $\Gamma^{\rm ch}$ yields a
submanifold $P(\Gamma^{\rm ch})$ of $P(\Gamma)$ whose boundary is
a lens space $L(p',q')$. The closure of $P(\Gamma)\setminus
P(\Gamma^{\rm ch})$ is a smooth compact 4-manifold  with oriented
boundary $-L(p',q')\bigsqcup\partial P(\Gamma)$ where $\bigsqcup$
denotes the disjoint sum operation. Starting with several chains
$\Gamma^{\rm ch}_s$ ($s=\overline{1,N}$) in $\Gamma$ (where
$\overline{0,N}$ is the integer numbers interval from 0 to $N$),
one can introduce a cobordism $P(\Gamma_D)$ between
$\Sigma(\Gamma):=\partial P(\Gamma)$ and the disjoint union
$L=\displaystyle \bigsqcup_{s=1}^N L(p'_s,q'_s)$, {\it i.e.} \be
\label{2.5} \partial P(\Gamma_D)=\left(-\bigsqcup_{s=1}^N
L(p'_s,q'_s)\right)\bigsqcup\Sigma(\Gamma). \ee Such a cobordism
will be called {\it graph cobordism}. Here naturally appears the
concept of a {\it decorated graph} $\Gamma_D$ shown as an ordinary
graph $\Gamma$ but with ovals or circles, each enclosing exactly
one chain $\Gamma^{\rm ch}_s$.

\noindent{\it Observation 2.1.} The chains $\left\{\Gamma^{\rm
ch}_s\right\}$ must be disjoint in the following sense: No two
chains should have a common vertex, and no edges of $\Gamma$
should have one endpoint on one chain and another, on any other
chain.

\noindent{\it Observation 2.2.} Consider the algebraic case when
all extra lens spaces $L(p'_k,q'_k)$ are subject to the condition
$p'_k<0$, and there are decorated (by ovals) $N_{\rm extra}$ extra
lens spaces and $N_{\rm leaf}$ leaf lens spaces $L(-a_i,b_i)\equiv
L(p'_i,q'_i)$. In this case the cobordism $P(\Gamma_D)$ has the
boundary \be \label{2.6} \partial P(\Gamma_D)=\left(-\bigsqcup_{k=
1 }^{N_{\rm extra}}L(p'_k,q'_k)\right)\left(-\bigsqcup_{i=1
}^{N_{\rm leaf}}L(p'_i,q'_i)\right)\bigsqcup\Sigma_{\rm alg}, \ee
and its intersection matrix $Q(\Gamma_D)$ is negative defined. The
series defining transition amplitudes (partition functions) in
section \ref{s3} converge if the intersection matrix of the
four-dimensional cobordism is positive defined. Hence the
orientation of the spacetime cobordism describing a topological
tunnelling should be inverse to that of $P(\Gamma_D)$. We define
the graph cobordism as $M_D=-P(\Gamma_D)=P(-\Gamma_D)$ where
$-\Gamma_D$ is the decorated graph $\Gamma_D$ with all weights
being sign-inverse. Thus \be \label{2.7} \partial
M_D=\left(-\bigsqcup_{k= 1 }^{N_{\rm extra}}L(|p'_k|,q'_k)\right)
\left(-\bigsqcup_{i=1 }^{N_{\rm leaf}}L(a_i,b_i)\right)\bigsqcup
\Sigma \ee where $\Sigma=\Sigma(-\Gamma)=-\Sigma_{\rm alg}$, and
it was taken into account that $p'_k<0$ and $a_i=-p'_i>0$. Since
$N=N_{\rm extra}+N_{\rm leaf}$, (\ref{2.7}) now reads \be
\label{2.8} \partial M_D=\left(-\bigsqcup_{s=1}^NL(|p'_s|,q'_s)
\right)\bigsqcup \Sigma. \ee The resulting orientation of the
four-manifold $M_D$ coincides with that introduced by Hirzebruch
in \cite{Hirz} for plumbed manifolds. The graph cobordisms $M_D$
have positive defined intersection forms. In section \ref{s4} we
build examples of such cobordisms as a spacetime basis for
cosmological models.

\subsection{Principal $U(1)$-bundles, connections and cohomologies}
\label{s2.2}

Let $M$ be a four-manifold with boundary $\partial M$. We denote
by $i_\partial : \partial M \rightarrow M$ the natural inclusion
map and by $d$, the de Rham differential on $M$. By $H^p(M,
\mathbb{Z})$ we further denote the absolute $p$th cohomology group
with integer coefficients, and by $H^p(M,\partial M,\mathbb{Z})$,
the relative $p$th cohomology group modulo $\partial M$. We also
denote by $\Omega^p(M)=C^p(M,\mathbb{R})$ the space of $p$-forms
on $M$, and by $\Omega^p(M,\partial M)=C^p(M,\partial
M,\mathbb{R})$ the space of relative $p$-forms on $M$. The
subscript $\mathbb{Z}$ is attached for corresponding subsets of
$p$-forms with integer (relative) periods: \be \label{2.9}
\int_\Sigma f\in\mathbb{Z} \ee where $f\in\Omega^p_\mathbb{Z}(M)$
(or $\Omega^p_\mathbb{Z}(M,\partial M)$, while $\Sigma$ is
(relative) closed $p$-dimensional surface (cycle).

The quantization of BF-theory and evaluation of the transition
amplitude corresponding to the cobordism $M$ involve summation
over the topological classes of gauge fields. Mathematically,
these classes can be identified with the isomorphism classes of
principal $U(1)$-bundles.

Let $\textnormal{Princ}\,(M)$ be the group of principal
$U(1)$-bundles over $M$. It is well known that there exists the
isomorphism \be \label{2.10} \textnormal{Princ}\,(M)
\overset{c}{\cong}H^2(M,\mathbb{Z}) \ee which assigns to a bundle
$P$ its first Chern class $c(P)$, see, {\it e.g.}, \cite{Zu1,
Zu2}. Since $H^2(M,\mathbb{Z})$ is an Abelian group, the subset of
elements of finite order is a subgroup, $\textnormal{Tor}
H^2(M,\mathbb{Z})$, called the torsion subgroup of
$H^2(M,\mathbb{Z})$. The preimage by the map $c$ of the torsion
subgroup is the subgroup $\textnormal{Princ}_{_0}(M)$ of
$\textnormal{Princ}\,(M)$, the elements of which are called flat
principal $U(1)$-bundles, {\it i.e.} \be \label{2.11}
\textnormal{Princ}_{_0}(M)\overset{c}{\cong}\textnormal{Tor}
H^2(M,\mathbb{Z}). \ee

A relative principal $U(1)$-bundle $(P,t)$ on $M$ consists both of
principal $U(1)$-bundle $P$ on $M$ such that its restriction
$i^*_\partial P$ on $\partial M$ is trivial, and of trivialization
$t:i^*_\partial P\rightarrow\partial M\times U(1)$. The relative
principal $U(1)$-bundles form a group $\textnormal{Princ}(M,
\partial M)$ which is isomorphic to the 2nd relative cohomology
group, \be \label{2.12} \textnormal{Princ}(M,
\partial M)\overset{c_{\rm rel}}{\cong}H^2(M,\partial M,\mathbb{Z}),
\ee where the map $c_{\rm rel}$ assigns to a relative bundle
$(P,t)$ its relative first Chern class $c(P,t)=c_{\rm rel}(P)$.

Of course, we can describe the group $\textnormal{Princ}(\partial
M)$ of principal $U(1)$-bundles on $\partial M$ in the same way as
we did for the group $\textnormal{Princ}(M)$. Thus the isomorphism
(\ref{2.10}) holds in the form \be \label{2.13} \textnormal{Princ}
(\partial M)\overset{c_\partial}{\cong}H^2(\partial M,\mathbb{Z}).
\ee The preimage of $\textnormal{Tor}H^2(\partial M,\mathbb{Z})$
relative to the map $c_\partial$ is the subgroup\linebreak
$\textnormal{Princ}_{_0}(\partial M)$ of flat principal
$U(1)$-bundles on $\partial M$.

Consider now the problem of extendability of principal bundles on
$\partial M$ to $M$ following Zucchini \cite{Zu2}. This is
important since the gauge theory transition amplitude (partition
function) involves a sum over the set of the bundles
$P\in\textnormal{Princ}(M)$ such that their restrictions
$i^*_\partial P$ on the boundary $\partial M$ coincide with a
fixed bundle $P_\partial\in\textnormal{Princ} (\partial M)$. Every
bundle $P\in\textnormal{Princ}(M)$ yields by pull-back
$i^*_\partial:P\rightarrow P_\partial$ (induced by the natural
inclusion $i_\partial:\partial M\rightarrow M$) a bundle
$P_\partial\in\textnormal{Princ}(\partial M)$. But the converse is
in general false: not every bundle $P_\partial$ is a pull-back of
some bundle $P$. When this does indeed happen, one says that
$P_\partial$ is extendable to $M$. We shall show that in the case
of four-dimensional graph cobordisms (considered in this paper as
a model of spacetime) any
$P_\partial\in\textnormal{Princ}(\partial M)$ is extendable. To
this end, consider the absolute/relative cohomology exact
sequence: \be \label{2.14} ...\rightarrow H^p (\partial
M,\mathbb{Z})\rightarrow H^{p+1}(M,\partial M,
\mathbb{Z})\rightarrow H^{p+1}(M,\mathbb{Z})\rightarrow H^{p+1}
(\partial M,\mathbb{Z})\rightarrow... . \ee We can now use the
isomorphisms (\ref{2.10}), (\ref{2.12}) and (\ref{2.13}) to draw
the commutative diagram \be \label{2.15} \begin{array}{ccccc} H^1
(\partial M,\mathbb{Z})\rightarrow &
\hspace*{-10.pt}H^2(M,\partial M, \mathbb{Z})
\overset{j^*}{\rightarrow} & \hspace*{-10.pt}H^2(M,\mathbb{Z})
\overset{i^*_\partial}{\rightarrow} & \hspace*{-10.pt}H^2(\partial
M,\mathbb{Z}) \overset{\delta^*}{\rightarrow} &
\hspace*{-9.pt}H^3(M,\partial
M,\mathbb{Z})\\
& \hspace*{-10.pt}c_{\rm rel}\uparrow & \hspace*{-10.pt}c\uparrow
& \hspace*{-10.pt}c_\partial\uparrow & \\
& \hspace*{-10.pt}\textnormal{Princ}(M,\partial
M)\overset{j^*}{\rightarrow} &
\hspace*{-10.pt}\textnormal{Princ}(M)\overset{i^*_\partial}{\rightarrow}
& \hspace*{-10.pt}\textnormal{Princ}(\partial M) &
\end{array} \ee in which the lines are exact and vertical maps are
isomorphisms. In the case of graph cobordisms under consideration
it is true that \cite{FFU, SavArt} \be \label{2.16} H^3(M,\partial
M,\mathbb{Z})=0, \ee \be \label{2.17} H^1 (\partial
M,\mathbb{Z})=0. \ee Interpretation of the second line in
(\ref{2.15}) is quite simple: the mapping $j^*$ associates with
every relative bundle $(P,t)\in \textnormal{Princ}(M,\partial M)$
the underlying bundle $P\in \textnormal{Princ}(M)$, and the
mapping $i^*_\partial$ associates with every bundle $P$ its
pull-back bundle $P_\partial=i^*_\partial
P\in\textnormal{Princ}(\partial M)$. This interpretation then
applies to the first line too.

By the exactness of lines in (\ref{2.15}) the bundle
$P_\partial\in\textnormal{Princ}(\partial M)$ is the pull-back of
the bundle $P\in \textnormal{Princ}(M)$ iff \be \label{2.18}
\delta^*(c_\partial(P_\partial))=0. \ee Consequently, the
obstruction to the extendability of $P_\partial$ is a class
of\linebreak $H^3(M,\partial M,\mathbb{Z})$. Since (\ref{2.16}) is
true for all our graph cobordisms, every principal $U(1)$-bundles
on $\partial M$ are extendable onto $M$. Each $P_\partial$ has
several extensions to $M$. Again, by the exactness of
(\ref{2.15}), its extensions are parametrized by the group of
relative bundles $\textnormal{Princ}(M,\partial M)$. This
parametrization is one-to-one due to (\ref{2.17}) (the mapping
$j^*$ is injective).

\subsection*{Connections}

Let $P\in\textnormal{Princ}(M)$ be a principal $U(1)$-bundle. We
denote by $\textnormal{Conn}(P)$ the affine space of connections
of $P$. Let $A\in\textnormal{Conn}(P)$ be a connection on $P$. We
can fix a trivializing cover $\{U_\alpha\}$ of $M$ ($M=\cup_\alpha
U_\alpha$) and assign to each open set $U_\alpha$ a vector
potential $A^\alpha$. The connection $\{A^\alpha\}$ is a \v{C}ech
0-cochain with values in 1-forms $A^\alpha=A^\alpha_idx^i$
\cite{OrlAlv}. The curvature $F_A$ of a connection $A$ is defined
by \be \label{2.19} F_A=dA. \ee This is a brief expression of the
local relations $F^\alpha_A=F_A|_{U_\alpha}=dA^\alpha$. The gauge
transformation properties of $A$ \cite{OlAl1} \be \label{2.20}
A^\alpha-A^\beta=d\chi^{\alpha\beta} \ee ensure that $F_A$ does
not depend on the chosen local trivialization of $P$, {\it i.e.}
$F$ is gauge invariant, $F^\alpha=F^\beta$ in $U_\alpha\cap
U_\beta$, thus $F_A\in\Omega^2(M)$ is a 2-form, $F_A$ obviously
being closed: \be \label{2.21} dF_A=0. \ee

\subsection{Intersection forms and the Poincar\'e--Lefschetz duality}
\label{s2.3}

Now let $M$ be a graph cobordism corresponding to a decorated
graph $\Gamma_D$ (see subsection \ref{s2.1} for $M_D$; we suppress
the subindex $_D$ below in this section). Then for any elements
$f,f'\in H^2(M, \mathbb{Z})$ the rational intersection number
$\langle f,f' \rangle_{\mathbb{Q}}$ is defined as follows
\cite{FFU}: We start with the part \be \label{2.22} 0\rightarrow
H^2(M,\partial M,\mathbb{Z})\overset{j^*}
{\rightarrow}H^2(M,\mathbb{Z})
\overset{i^*_\partial}{\rightarrow}H^2(\partial M,\mathbb{Z})
\rightarrow 0 \ee of the exact sequence (\ref{2.14}). Since
$H^2(\partial M,\mathbb{Z})$ is a pure torsion (finite Abelian
group), we see that for any $f\in H^2(M, \mathbb{Z})$ there exists
$p\in\mathbb{Z}$ with $i^*_\partial(pf) =0$, hence $pf=j^*(b)$ for
unique $b\in H^2(M,\partial M,\mathbb{Z})$. Then we put \be
\label{2.23} \langle f,f' \rangle_{\mathbb{Q}}:=\frac{1}{p}\langle
b,f' \rangle_{\mathbb{Z}} \in \mathbb{Q}, \ee where $\langle ~ ~,
~ ~\rangle_{\mathbb{Z}}$ on the right-hand side is the usual
integer intersection number well defined due to the
Poincar\'e--Lefschetz duality \cite{SavB2} \be \label{2.24}
H^2(M,\mathbb{Z})\cong H_2(M,\partial M, \mathbb{Z}), \ee \be
\label{2.25} H^2(M,\partial M,\mathbb{Z}) \cong H_2(M,\mathbb{Z}).
\ee The Poincar\'e--Lefschetz duality pairing (PL-pairing) \be
\label{2.26} \langle ~ ~, ~ ~\rangle_{\mathbb{Z}}:H^2(M,\partial
M,\mathbb{Z})\times H^2(M,\mathbb{Z})\rightarrow\mathbb{Z} \ee can
be written in the de Rham representation as \be \label{2.27}
\langle b,f'\rangle= \int_Mb\wedge f' \in\mathbb{Z}. \ee This is
true since $\Lambda=H^2(M,\partial M,\mathbb{Z})$ and
$\Lambda^\#=H^2(M,\mathbb{Z})$ are the integer cohomology lattices
in the de Rham cohomology space $H^2(M, \mathbb{R})$. Note that if
$H^2(M,\partial M,\mathbb{Z})$ and $H^2(M,\mathbb{Z})$ had
torsion, this inclusion would be impossible, but in the case of
graph cobordisms these groups are finitely generated free Abelian
ones \cite{FFU, NemNicol}. Moreover,
$$
\textnormal{rank}\,H^2(M,\partial M,\mathbb{Z})=
\textnormal{rank}\, H^2(M,\mathbb{Z})=\textnormal{rank}\,H^2(M,
\mathbb{R})
$$
since $H^2(\partial M,\mathbb{Z})$ is pure torsion \cite{HiWy}.

From exactness of the sequence (\ref{2.22}) it follows that
$H^2(M,\partial M,\mathbb{Z})$ is the subgroup of $H^2(M,
\mathbb{Z})$ (the mapping $j^*$ is monomorphism). Since both
groups are torsion-free, the group $H^2(M,\mathbb{Z})$ can be
represented as a homomorphism group \be \label{2.28}
H^2(M,\mathbb{Z})=\textnormal{Hom}(H_2(M,\mathbb{Z}),\mathbb{Z})
\cong\textnormal{Hom}(H^2(M,\partial M,\mathbb{Z}),\mathbb{Z}),
\ee {\it i.e.} as the lattice $\Lambda^\#$ dual to $\Lambda=
H^2(M,\partial M,\mathbb{Z})$ with respect to the scalar product
(PL-paiting) \be \label{2.29} \Lambda^\#=H^2(M,\mathbb{Z}):=
\left\{f\in H^2(M,\mathbb{R})|\langle b,f\rangle\in\mathbb{Z},
\forall b\in\Lambda\right\} \ee where $\langle b,f\rangle$ is
defined in de Rhamian representation by (\ref{2.27}). Note that
$\Lambda\subset\Lambda^\#$, thus $\Lambda$ is not unimodular, and
it is possible to introduce the nontrivial discriminant group \be
\label{2.30} T(\Lambda):= \Lambda^\#/\Lambda=H^2(M,\mathbb{Z})/
H^2(M,\partial M,\mathbb{Z})=H^2(\partial M,\mathbb{Z}) \ee which
is the finite Abelian group. (The last equality follows from
exactness of the sequence (\ref{2.22})).

For our purposes there will be useful the following\newline {\it
Proposition} \cite{FFU}. Let $M$ be a graph cobordism. If we
choose a certain basis $b_I$ of $\Lambda= H^2(M,\partial
M,\mathbb{Z})$ and the dual basis $f^I$ of
$\Lambda^\#=H^2(M,\mathbb{Z})$ ($I=1,...,r= \textnormal{rank}\,
H^2(M,\mathbb{Z})$), dual in the sense that $\langle
b_I,f^J\rangle_\mathbb{Z}=\delta^J_I$, then the integral
intersection matrix \be \label{2.31} Q_{IJ}=\langle
b_I,b_J\rangle_\mathbb{Z} \ee for $H^2(M,\partial M,\mathbb{Z})$,
is inverse of the rational intersection matrix \be \label{2.32}
Q^{IJ}=\langle f^I,f^J\rangle_\mathbb{Q}. \ee
Note that order of the discriminant group (\ref{2.30}) is
\cite{OlAl2} \be \label{2.33} |T(\Lambda)|=\textnormal{det}\,
\langle b_I,b_J\rangle_\mathbb{Z}=\textnormal{det}\,Q_{IJ}. \ee

Let us now calculate the discriminant group $T(\Lambda)$ defined
in (\ref{2.30}). The exactness of the cohomology sequence
(\ref{2.22}) results in existence of such a basis
$\left\{f^I\left|I\in\overline{1,r}\right.\right\}$ in the group
$H^2(M,\mathbb{Z})$ that $i^*_\partial(f^I)=t^I$ are generators of
$H^2(\partial M,\mathbb{Z})$ \cite{FintSt}. Due to finiteness of
the group $T(\Lambda)=H^2(\partial M,\mathbb{Z})$ there exist
minimal integers $p(I)>1$ such that $p(I)t^I=0$ (without summation
in $I$). Since the mapping $i^*_\partial$ is linear, $i^*_\partial
(p(I)f^I)=p(I)t^I=0$. Moreover, the monomorphism property of $j^*$
in (\ref{2.22}) yields existence of the unique element
$\tilde{f}_I\in H^2(M,\partial M,\mathbb{Z})$ such that
$j^*(\tilde{f}_I)=p(I)f^I$. The class $j^*(\tilde{f}_I)$ can be
considered as an element $\tilde{f}^I$ in the subgroup
$H^2(M,\partial M,\mathbb{Z})$ of $H^2(M,\mathbb{Z})$. Due to
$\textnormal{rank}\,H^2(M,\partial M,\mathbb{Z})=
\textnormal{rank}\, H^2(M,\mathbb{Z})=r$ and to the minimality of
the integers $p(I)$, the set of classes
$\left\{\tilde{f}^I\right\}= \left\{p(I)f^I\right\}$ forms the
basis in $H^2(M,\partial M,\mathbb{Z})$ \cite{FFU}. The theorems
{\bf 5.1.1} and {\bf 5.1.3} in \cite{HiWy} then lead to the
conclusion that the discriminant group $T(\Lambda)$ reads as \be
\label{2.34} T(\Lambda)=\bigoplus^r_{I=1}\mathbb{Z}_{p(I)}. \ee
Orders $p(I)$ of cyclic groups $\mathbb{Z}_{p(I)}$ can be
calculated from the characteristics of the decorated graph
$\Gamma_D$ which determines topology of the cobordism $M$, {\it
i.e.} from the topological invariants of this graph cobordism. To
this end note that the number of elements in the discriminant
group $T(\Lambda)$ is \be \label{2.35} |T(\Lambda)|=p(1)\cdots
p(r)=\textnormal{det}Q_{IJ}=|p'_1\cdots p'_{2r+1}|. \ee The last
equality follows from juxtaposition of the generators
$\left\{f^I\left|I\in\overline{1,r}\right.\right\}$ of
$H^2(M,\mathbb{Z})$ to the vertices
$\left\{v_I\left|I\in\overline{1,r}\right.\right\}$ of the graph
$\Gamma_D$ outside of all the decorated ovals \cite{SavArt,
NeumWahl}. In this case \be \label{2.36}
\textnormal{det}Q_{IJ}=|p'_1\cdots p'_{2r+1}| \ee where $p'_s$ are
characteristics of decorated ovals (decorated linear chains) of
the graph $\Gamma_D$, see (\ref{2.8}) with $N=2r+1$ and the first
decorated graph $\Gamma^r_D$ in figure 8.

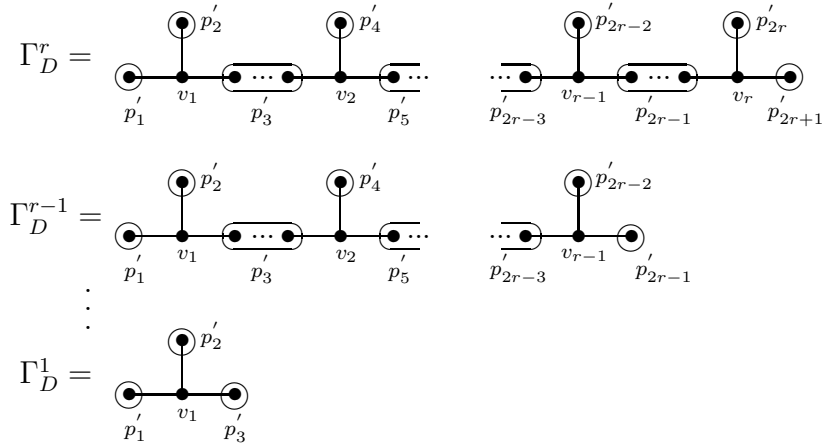
\begin{figure}[h]
\begin{center}
\setlength{\unitlength}{1pt}
\begin{picture}(150,50)

\put(-90,5){\makebox(5,5){$\Gamma^{r}_{D}=$}}
\put(-60,-15){\makebox(5,5){\scriptsize $p^{'}_{1}$}}
\put(-31,20){\makebox(5,5){\scriptsize $p^{'}_{2}$}}
\put(-12,-15){\makebox(5,5){\scriptsize $p^{'}_{3}$}}
\put(29,20){\makebox(5,5){\scriptsize $p^{'}_{4}$}}
\put(40,-15){\makebox(5,5){\scriptsize $p^{'}_{5}$}}
\put(125,20){\makebox(5,5){\scriptsize $p^{'}_{2r-2}$}}
\put(85,-15){\makebox(5,5){\scriptsize $p^{'}_{2r-3}$}}
\put(181,20){\makebox(5,5){\scriptsize $p^{'}_{2r}$}}
\put(140,-15){\makebox(5,5){\scriptsize $p^{'}_{2r-1}$}}
\put(190,-15){\makebox(5,5){\scriptsize $p^{'}_{2r+1}$}}
\put(-40,-10){\makebox(5,5){\scriptsize $v_{1}$}}
\put(20,-10){\makebox(5,5){\scriptsize $v_{2}$}}
\put(110,-10){\makebox(5,5){\scriptsize $v_{r-1}$}}
\put(170,-10){\makebox(5,5){\scriptsize $v_{r}$}}
\put(-60,0){\line(1,0){40}}
\put(-40,0){\line(0,1){20}}
\put(-60,-.5){\makebox(.5,.5){$\bullet$}}
\put(-20,-.5){\makebox(.5,.5){$\bullet$}}
\put(-40,20){\makebox(.5,.5){$\bullet$}}
\put(-40,-.5){\makebox(.5,.5){$\bullet$}}
\put(-60,0){\circle{10}}
\put(-40,20){\circle{10}}
\multiput(-14,-.5)(3,0){3}{.}
\put(-9.5,0){\oval(30,10)}
\put(0,0){\line(1,0){40}}
\put(20,0){\line(0,1){20}}
\put(0,-.5){\makebox(.5,.5){$\bullet$}}
\put(40,-.5){\makebox(.5,.5){$\bullet$}}
\put(20,20){\makebox(.5,.5){$\bullet$}}
\put(20,-.5){\makebox(.5,.5){$\bullet$}}
\put(20,20){\circle{10}}
\multiput(45,-.5)(3,0){3}{.}
\multiput(76,-.5)(3,0){3}{.}
\put(50,0){\oval(30,10)[l]}
\put(81,0){\oval(30,10)[r]}
\put(90,0){\line(1,0){40}}
\put(110,0){\line(0,1){20}}
\put(90,-.5){\makebox(.5,.5){$\bullet$}}
\put(130,-.5){\makebox(.5,.5){$\bullet$}}
\put(110,20){\makebox(.5,.5){$\bullet$}}
\put(110,-.5){\makebox(.5,.5){$\bullet$}}
\put(110,20){\circle{10}}
\put(140,0){\oval(30,10)}
\multiput(136,-.5)(3,0){3}{.}
\put(150,0){\line(1,0){40}}
\put(170,0){\line(0,1){20}}
\put(150,-.5){\makebox(.5,.5){$\bullet$}}
\put(190,-.5){\makebox(.5,.5){$\bullet$}}
\put(170,20){\makebox(.5,.5){$\bullet$}}
\put(170,-.5){\makebox(.5,.5){$\bullet$}}
\put(170,20){\circle{10}}
\put(190,0){\circle{10}}

\put(-90,-55){\makebox(5,5){$\Gamma^{r-1}_{D}=$}}
\put(-60,-75){\makebox(5,5){\scriptsize $p^{'}_{1}$}}
\put(-31,-40){\makebox(5,5){\scriptsize $p^{'}_{2}$}}
\put(-12,-75){\makebox(5,5){\scriptsize $p^{'}_{3}$}}
\put(29,-40){\makebox(5,5){\scriptsize $p^{'}_{4}$}}
\put(40,-75){\makebox(5,5){\scriptsize $p^{'}_{5}$}}
\put(125,-40){\makebox(5,5){\scriptsize $p^{'}_{2r-2}$}}
\put(85,-75){\makebox(5,5){\scriptsize $p^{'}_{2r-3}$}}
\put(140,-75){\makebox(5,5){\scriptsize $p^{'}_{2r-1}$}}
\put(-40,-70){\makebox(5,5){\scriptsize $v_{1}$}}
\put(20,-70){\makebox(5,5){\scriptsize $v_{2}$}}
\put(110,-70){\makebox(5,5){\scriptsize $v_{r-1}$}}
\put(-60,-60){\line(1,0){40}}
\put(-40,-60){\line(0,1){20}}
\put(-60,-60.5){\makebox(.5,.5){$\bullet$}}
\put(-20,-60.5){\makebox(.5,.5){$\bullet$}}
\put(-40,-40){\makebox(.5,.5){$\bullet$}}
\put(-40,-60.5){\makebox(.5,.5){$\bullet$}}
\put(-60,-60){\circle{10}}
\put(-40,-40){\circle{10}}
\multiput(-14,-60.5)(3,0){3}{.}
\put(-9.5,-60){\oval(30,10)}
\put(0,-60){\line(1,0){40}}
\put(20,-60){\line(0,1){20}}
\put(0,-60.5){\makebox(.5,.5){$\bullet$}}
\put(40,-60.5){\makebox(.5,.5){$\bullet$}}
\put(20,-40){\makebox(.5,.5){$\bullet$}}
\put(20,-60.5){\makebox(.5,.5){$\bullet$}}
\put(20,-40){\circle{10}}
\multiput(45,-60.5)(3,0){3}{.}
\multiput(76,-60.5)(3,0){3}{.}
\put(50,-60){\oval(30,10)[l]}
\put(81,-60){\oval(30,10)[r]}
\put(90,-60){\line(1,0){40}}
\put(110,-60){\line(0,1){20}}
\put(90,-60.5){\makebox(.5,.5){$\bullet$}}
\put(130,-60.5){\makebox(.5,.5){$\bullet$}}
\put(110,-40){\makebox(.5,.5){$\bullet$}}
\put(110,-60.5){\makebox(.5,.5){$\bullet$}}
\put(110,-40){\circle{10}}
\put(130,-60.5){\circle{10}}

\multiput(-77,-95)(0,7){3}{.}

\put(-90,-115){\makebox(5,5){$\Gamma^{1}_{D}=$}}
\put(-60,-135){\makebox(5,5){\scriptsize $p^{'}_{1}$}}
\put(-31,-100){\makebox(5,5){\scriptsize $p^{'}_{2}$}}
\put(-22,-135){\makebox(5,5){\scriptsize $p^{'}_{3}$}}
\put(-40,-130){\makebox(5,5){\scriptsize $v_{1}$}}
\put(-60,-120){\line(1,0){40}}
\put(-40,-120){\line(0,1){20}}
\put(-60,-120.5){\makebox(.5,.5){$\bullet$}}
\put(-20,-120.5){\makebox(.5,.5){$\bullet$}}
\put(-40,-100){\makebox(.5,.5){$\bullet$}}
\put(-40,-120.5){\makebox(.5,.5){$\bullet$}}
\put(-60,-120){\circle{10}}
\put(-40,-100){\circle{10}}
\put(-20,-120.5){\circle{10}}

\end{picture}\\
\vspace*{5.cm} \caption{The decorated
graphs $\Gamma^{I}_{D}$.}
\end{center}
\end{figure}

Now consider sublattices $\Lambda_{r-1}$ and $\Lambda^\#_{r-1}$
(of the lattices $\Lambda_r$ and $\Lambda^\#_r$) generated by the
subbases
$\left\{\tilde{f}^I\left|I\in\overline{1,r-1}\right.\right\}$ and
$\left\{f^I\left|I\in\overline{1,r-1}\right.\right\}$,
respectively. This singles out from the graph
$\Gamma_D\equiv\Gamma^r_D$, the subgraph $\Gamma^{r-1}_D$ which
consists of the vertices
$\left\{v_I\left|I\in\overline{1,r-1}\right.\right\}$ and
decorated ovals adjusted to these vertices, {\it i.e.} decorated
chains with characteristics $p'_1,...,p'_{2(r-1)+1}=p'_{2r-1}$,
see figure 8. The discriminant group corresponding to this
subgraph is \be \label{2.37}
T(\Lambda_{r-1})=\Lambda^\#_{r-1}/\Lambda_{r-1}=
\bigoplus^{r-1}_{I=1}\mathbb{Z}_{p(I)} \ee and possesses the order
\be \label{2.38} |T(\Lambda_{r-1})|=p(1)\cdots p(r-1)=|p'_1\cdots
p'_{2r-1}|. \ee A comparison of (\ref{2.38}) and (\ref{2.35})
shows that \be \label{2.39} p(r)=|p'_{2r}p'_{2r+1}|. \ee After a
finite number of steps, we encounter the following sequence of
relations: \be \label{2.40} p(I)=|p'_{2I}p'_{2I+1}|\textnormal{
for }2\leq I\leq r\textnormal{ and finally }p(1)=|p'_1p'_2p'_3|
\ee (the last expression is related to the final graph
$\Gamma^1_D$).

\section{Topological gauge theory} \label{s3} \setcounter{equation}{0}

\subsection{The classical Abelian BF-model} \label{s3.1}

Let $P\in\textnormal{Princ}\,(M)$ be a principal $U(1)$-bundle on
$M$ and $P_\partial:=i^*_\partial P\in\textnormal{Princ}\,
(\partial M)$ be the induced principal $U(1)$-bundle on $\partial
M$. The Abelian BF gauge theory action is a functional of the
connection $A\in\textnormal{Conn}\,(P)$ and of the auxiliary
2-cochain $B\in C^2(M,\mathbb{R})$ (2-form in the de Rham
representation) \cite{Hor} \be \label{3.1} S=\frac{1}{2\pi} \int_M
\left(B\wedge F-\frac{\lambda}{2}B\wedge B\right) \ee where $F=dA$
is the curvature of the connection $A$ (see subsection
\ref{s2.2}), $\lambda$ being a scale factor analogous to the
cosmological constant \cite{BaezMG}.

The dynamical equations following from (\ref{3.1}) are quite
simple, \be \label{3.2} F=\lambda B, \ee \be \label{3.3} dB=0, \ee
if the normal boundary condition \cite{Zu2} on the variation of
the connection $\delta A$ is accepted, \be \label{3.4}
i^*_\partial(\delta A)=0. \ee Moreover the equation (\ref{3.3}) is
a consequence of (\ref{3.2}) and the Bianchi identity $dF=0$. The
action (\ref{3.1}) is actually invariant under very large gauge
transformations \be \label{3.5} \delta A=w, \ee \be \label{3.6}
\delta B=\frac{1}{\lambda}dw \ee where $w$ is an arbitrary 1-form.
The gauge-inequivalent classical solutions of the BF-system
(\ref{3.1}) are thus characterized by a 2-cocycle $F_{\rm cl}$
modulo coboundary $dw$ ({\it i.e.} $F_{\rm cl}\in
H^2(M,\mathbb{R})$) and by a 2-cocycle $B_{\rm cl}$ modulo
coboundary $\frac{1}{\lambda}dw$ ({\it i.e.} $B_{\rm cl}\in
H^2(M,\mathbb{R})$).

Quantization of the BF-theory involves a summation over
topological classes of gauge fields. Formally, these classes may
be identified with the isomorphism classes of principal
$U(1)$-bundles $\textnormal{Princ}\,(M)$ and of relative principal
$U(1)$-bundles $\textnormal{Princ}\,(M,\partial M)$ which were
described in subsection \ref{s2.1} by means of isomorphisms
(\ref{2.10}) and (\ref{2.12}). Thus it is natural to introduce the
BF analogue of the generalized Dirac quantization conditions
({\it cf.} \cite{Verl}) for a graph cobordism $M$ as follows: \be
\label{3.7} c_{\rm rel}(P)=\frac{B_{\rm cl}}{2\pi}\in\Lambda\cong
H^2(M,\partial M,\mathbb{Z})\subset H^2(M,\mathbb{R}), \ee \be
\label{3.8} c(P)=\frac{F_{\rm cl}}{2\pi}\in\Lambda^\#\cong H^2(M,
\mathbb{Z})\subset H^2(M,\mathbb{R}). \ee

\subsection{Transition amplitudes} \label{s3.2}

In an analogy with the partition function of Abelian gauge theory
\cite{Verl, WitS} we construct the transition amplitude for purely
topological gauge (BF-) theory over a graph cobordism $M$. As
usual \cite{Zu2}, this transition function involves both summation
over the set of principal $U(1)$ bundles
$P\in\textnormal{Princ}(M)\cong H^2(M,\mathbb{Z})$ such that the
principal bundle $P_\partial=i^*_\partial P$ is fixed, and
functional integration over bulk quantum fluctuations $v$ of the
connection on $P$ which satisfies the ordinary normal boundary
condition \cite{WittSL} \be \label{3.9} i^*_\partial v=0. \ee We
apply the customary classical-background-quantum-splitting method
\cite{Dij} which can be realized as follows. We write a general
representative of cohomology class in $H^2(M,\mathbb{Z})$ as \be
\label{3.10} \frac{F}{2\pi}=n_If^I+dv=m^Ib_I+l_If^I+dv \ee (a
cohomology class and its representative we usually denote by the
same symbol). This expression needs a more detailed comments: $v$
is a proper 1-form describing quantum fluctuations. The first two
terms of the last right-hand-side corresponding to non-trivial
cohomology classes, describe classical (background) solution of
the BF-theory, \be \label{3.11} \frac{F_{\rm cl}}{2\pi}=n_If^I=
m^Ib_I+l_If^I. \ee In this formula $\{b_I\}$ is basis of lattice
(group) $\Lambda=H^2(M,\partial M,\mathbb{Z})$ while $\{f^I\}$,
basis of the dual lattice (group) $\Lambda^\#=H^2(M,\mathbb{Z})$,
see subsection \ref{s2.3}. Since $\Lambda$ is a sublattice of
$\Lambda^\#$ and since $i^*_\partial(f^I)=t^I$ are generators of
the subgroup $\mathbb{Z}_{p(I)}$ of the discriminant group
$T(\Lambda)=\Lambda^\#/\Lambda=H^2(M,\mathbb{Z})/ H^2(M,\partial
M,\mathbb{Z})$, an arbitrary element of $\Lambda^\#$ can be
represented as a sum of some elements $m^Ib_I$ of the lattice
$\Lambda$ ($m^I\in\mathbb{Z}$) and of such a linear combination
$l_If^I$, so that $l_I \in\overline{0,p(I)-1}$. The restriction on
the values of the coefficients $l_I$ is due to $p(I)f^I\in\Lambda$
(there is no summation in $I$).

Now note that $i^*_\partial(l_If^I)=l_It^I\in H^2(\partial M,
\mathbb{Z})$, then due to the isomorphism (\ref{2.13}) the set of
numbers $l_I \in\overline{0,p(I)-1}$ determines the bundle
$P_\partial$ on the boundary $\partial M$ which should be fixed in
the calculation of a transition amplitude. Thus in this
calculation the summation has to be done over the sets
$m^I\in\mathbb{Z}$, {\it i.e.} the summation over
$P\in\textnormal{Princ}(M)\cong H^2(M,\mathbb{Z})$ reduces to that
over $P\in\textnormal{Princ} (M,\partial M)\cong H^2(M,\partial
M,\mathbb{Z})$ which coincides with the result in \cite{Zu2}.

Passing to the procedure of calculation of the transition
amplitude we see that the equation of motion for $B$ (\ref{3.2})
is in fact an algebraic constraint which we can substitute back to
(\ref{3.1}) in order to obtain a more usual form of the BF-action
\cite{Thom} \be \label{3.12} S=\frac{1}{4\pi\lambda}\int_MF\wedge
F. \ee Inserting the expression (\ref{3.10}) into (\ref{3.12}), we
find \be \label{3.13}
S(\lambda,\Bar{m},\underline{l})=\frac{\pi}{\lambda}
\left(m^I+Q^{IJ}l_J\right)Q_{IK}\left(m^K+Q^{KL}l_L\right) \ee
where $\Bar{m}=\left\{m^I|I\in\overline{1,r}\right\}$,
$\underline{l}=\left\{l_I|I\in\overline{1,r}\right\}$, and
$Q_{IJ}= \langle b_I,b_J\rangle_\mathbb{Z}=\int_Mb_I\wedge b_J$,
$Q^{IJ}=\langle f^I,f^J\rangle_\mathbb{Q}=\int_Mf^I\wedge f^J$ are
integer and rational intersection matrices defined in subsection
\ref{s2.3}; remember that $Q_{IJ}Q^{JK}=\delta^K_I$. Note that the
quantum fluctuations $v$ give zero contributions in (\ref{3.13})
due to the Stokes theorem \be \label{3.14} \int_Mdv\wedge dv=
\int_{\partial M}v\wedge dv=0 \ee and to the normal boundary
conditions (\ref{3.9}). Thus we see that the transition amplitude
(partition function corresponding to the graph cobordism $M$ in
Euclidean regime) reads
$$
Z(\lambda,\underline{l})=\frac{1}{c}\sum_{m^I\in\mathbb{Z}}\exp[
-S(\lambda,\Bar{m},\underline{l})]
$$
where the constant $c=1$ due to (\ref{3.14}), so that \be
\label{3.15} Z(\lambda,\underline{l})=\sum_{m^I\in\mathbb{Z}}\exp
\left[-\frac{\pi}{\lambda}\left(m^I+Q^{IJ}l_J\right)Q_{IK}\left(m^K
+Q^{KL}l_L\right)\right]. \ee This is recognizable as a sort of
theta function associated with the flux lattice
$\Lambda+Q\underline l$ since $m^Ib_I\in\Lambda$. To clarify this
question, we define the analogues of electric and magnetic fluxes
related to the field strength $\frac{F_{\rm cl}}{2\pi}\in
H^2(M,\mathbb{Z})$ through a set of homologically non-trivial
2-cycles. First note that in the case of a graph cobordism $M$
both integer intersection form (\ref{2.26}) and rational
intersection form (\ref{2.23}) are rigorously determined as  \be
\label{3.16} \langle ~ ~, ~
~\rangle_{\mathbb{Z}}:H^2(M,\mathbb{Z})\times H^2(M,\partial
M,\mathbb{Z})\rightarrow\mathbb{Z} \ee and \be \label{3.17}
\langle ~ ~, ~ ~\rangle_{\mathbb{Q}}:H^2(M,\mathbb{Z})\times
H^2(M,\mathbb{Z})\rightarrow\mathbb{Q}, \ee respectively. Due to
the Poincar\'e--Lefschetz duality in the form (\ref{2.24}) and
(\ref{2.25}), there are two induced pairings \be \label{3.18}
\langle ~ ~, ~ ~\rangle_{\mathbb{Z}}:H^2(M,\mathbb{Z})\times
H_2(M,\mathbb{Z})\rightarrow\mathbb{Z} \ee and \be \label{3.19}
\langle ~ ~, ~ ~\rangle_{\mathbb{Q}}:H^2(M,\mathbb{Z})\times
H_2(M,\partial M,\mathbb{Z})\rightarrow\mathbb{Q} \ee which enable
one to determine the fluxes of the field strength $\frac{F_{\rm
cl}}{2\pi}$ through non-trivial absolute and relative 2-cycles,
respectively.

Let us introduce a basis $\left\{\phi_I|I\in\overline{1,r}\right\}
$ of homologically non-trivial 2-cycles in $\Lambda\cong H_2(M,
\mathbb{Z})$ dual to the basis $\left\{f^I\right\}$ in
$\Lambda^\#\cong H^2(M,\mathbb{Z})$ with respect to the pairing
(\ref{3.18}) in the sense that \be \label{3.20} \langle f^I,\phi_J
\rangle_{\mathbb{Z}}=\delta^I_J. \ee Moreover, using once again
the Poincar\'e--Lefschetz duality (\ref{2.24}), we come from
(\ref{3.18}) to another induced pairing \be \label{3.21} \langle ~
~, ~ ~\rangle^{\rm hom}_{\mathbb{Z}}:H_2(M,\partial M,\mathbb{Z})
\times H_2(M,\mathbb{Z})\rightarrow\mathbb{Z} \ee which gives the
usual intersection numbers between the 2-cycles of\linebreak
$H_2(M,\partial M,\mathbb{Z})$ and $H_2(M,\mathbb{Z})$ \cite{FFU}.
Now, we can define a basis
$\left\{\beta^I|I\in\overline{1,r}\right\} $ of homologically
non-trivial relative 2-cycles in $\Lambda^\#\cong H_2(M,\partial
M,\mathbb{Z})$ dual to the basis $\left\{\phi_I \right\}$ with
respect to the pairing (\ref{3.21}) in the sense that \be
\label{3.22} \langle \beta^I,\phi_J \rangle^{\rm
hom}_{\mathbb{Z}}=\delta^I_J, \ee {\it i.e.} dual in the sense of
Poincar\'e--Lefschetz. In the definition of analogues of the
electric and magnetic fluxes we apply the Poincar\'e--Lefschetz
duality instead of the Hodge duality used in the ordinary
electrodynamics with a theta term \cite{Verl, WitS, OlAl2, OlAl3}.

We define the ``electric'' fluxes of field strength as fluxes
through the homologically non-trivial 2-cycles $\phi_I\in\Lambda$
using the scalar product (\ref{3.18}) as \be \label{3.23}
\Phi^{\rm(el)}_I(\Bar{m},\underline{l}):=\left\langle\frac{F_{\rm
cl}}{2\pi},\phi_I\right\rangle_{\mathbb{Z}}=Q_{IJ}m^J+l_I\in
\mathbb{Z}. \ee We analogously define ``magnetic'' fluxes of field
strength as fluxes through the homologically non-trivial relative
2-cycles $\beta^I\in\Lambda^\#$ using the scalar product
(\ref{3.19}) as \be \label{3.24} \Phi_{\rm(mag)}^I(\Bar{m},
\underline{l}):=\left\langle\frac{F_{\rm
cl}}{2\pi},\beta^I\right\rangle_{\mathbb{Q}}=m^I+Q^{IJ}l_J\in
\mathbb{Q}. \ee The expressions of the fluxes in terms of
``quantum numbers'' $m^I$ and $l_I$ in formulae (\ref{3.23}) and
(\ref{3.24}) follow from the relation (\ref{3.11}) and from the
duality of corresponding bases. Thus the transition amplitude
(\ref{3.15}) can be rewritten as \be \label{3.25}
Z(\lambda,\underline{l})=\sum_{\Bar{m}\in\Lambda}\exp
\left[-\frac{\pi}{\lambda}\Phi_{\rm(mag)}^I(\Bar{m},
\underline{l})Q_{IK}\Phi_{\rm(mag)}^K(\Bar{m},
\underline{l})\right]. \ee Consequently, this transition amplitude
describes the strong coupling of ``magnetic'' fluxes (\ref{3.24})
since the product $\frac{1}{\lambda}Q_{IJ}$ plays the r\^ole of
coupling constants' matrix. Here $Q_{IJ}$ is the integer
intersection matrix of the cobordism $M$ (with all non-zero
elements being $>1$), and the scale factor
$\frac{1}{\lambda}\geq1$ (to be shown in subsection \ref{s3.3}).

\noindent {\it Observation 3.1.} Note that if the lattice
$\Lambda$ is self-dual ($\Lambda\cong\Lambda^\#$), the
``electric'' and ``magnetic'' fluxes (in the sense of our
definition) mutually coincide, thus one has to introduce the Hodge
operator, which presumes existence of metric (and we are trying to
avoid this), to distinguish between these two types of fluxes.
Thus in the consideration of a closed 4-manifold ($\partial M=0$)
our model becomes trivial, and the non-triviality of our approach
is due to the substantiveness of the exact cohomological sequence
(\ref{2.22}) which degenerates into isomorphism,
$$
0\rightarrow H^2(M,\partial M,\mathbb{Z})\overset{j^*}
{\cong}H^2(M,\mathbb{Z})\rightarrow 0,
$$
if $\partial M=0$ or even $H^2(\partial M,\mathbb{Z})=0$.

To determine the behaviour of $Z(\lambda,\underline{l})$ under the
transformation $\lambda\rightarrow1/\lambda$ we use the same trick
as Olive and Alvarez \cite{OlAl2}. Using the Poisson summation
formula in matrix form \cite{GSW}
$$ \begin{array}{l}
\displaystyle\sum_{\Bar{m}\in\Lambda}\exp
\left[-\pi(\Bar{m}+\Bar{x})\cdot
A \cdot(\Bar{m}+\Bar{x})\right]=\\ ~ \\
(\textnormal{det}\,A)^{-1/2}
\displaystyle\sum_{\underline{n}\in\Lambda^\#}\exp
\left[-\pi\underline{n}\cdot A^{-1}\cdot\underline{n}+2\pi
i\underline{n}\cdot\Bar{x}\right],
\end{array}
$$
we can rewrite the transition amplitude (\ref{3.15}) in terms of
the weak coupling between ``electric'' fluxes (\ref{3.23}) as \be
\label{3.26} \left.\begin{array}{l}
Z(\lambda,\underline{l})=\lambda^{r/2}
(\textnormal{det}\,Q_{IJ})^{-1/2}\times\\
\displaystyle\sum_{n_I\in\mathbb{Z}}\exp
\left[\pi\lambda\left(2in_IQ^{IJ}l_J-n_IQ^{IJ}n_J\right)\right]=\\ ~ \\
\lambda^{r/2} (\textnormal{det}\,Q_{IJ})^{-1/2}
\displaystyle\sum_{\underline{l}'\in T(\Lambda)}\exp\left(2\pi
i\lambda l'_I Q^{IJ}l_J\right)\times\\
\displaystyle\sum_{\Bar{m}\in \Lambda}\exp\left[-\pi\lambda
\left(Q_{IJ}m^J+l'_I\right)Q^{IK}\left(Q_{KL}m^L+l'_K\right)
\right]=\\ ~ \\ \lambda^{r/2} (\textnormal{det}\,Q_{IJ})^{-1/2}
\displaystyle\sum_{\underline{l}'\in T(\Lambda)}\exp\left(2\pi
i\lambda l'_I Q^{IJ}l_J\right)\times\\
\displaystyle\sum_{\Bar{m}\in
\Lambda}\exp\left[-\pi\lambda\Phi^{\rm(el)}_I(\Bar{m},
\underline{l'})Q^{IK}\Phi^{\rm(el)}_K(\Bar{m},\underline{l'})
\right].
\end{array}\right\}\ee In the second and third parts of this
formula we used the expressions (\ref{3.11}) in the form \be
\label{3.27} \frac{F_{\rm cl}}{2\pi}=n_If^I= m^Ib_I+l'_If^I=
\left(Q_{IJ}m^J+l'_I\right)f^I=\Phi^{\rm(el)}_I(\Bar{m},
\underline{l'})f^I \ee and $b_I=Q_{IJ}f^J$ \cite{FFU}. The prime
in $l'_I$ (with respect to which the summation is performed)
distinguishes it from the fixed $l_I$, while the symbolic notation
$\underline{l}'\in T(\Lambda)$ means that the summation runs over
all collections $\{l'_I\}$ which determine the elements $l'_It^I$
of the discriminant group $T(\Lambda)$, {\it i.e.} $l'_I\in
\overline{1,p(I)-1}$.

The transition amplitudes $Z(\lambda,\underline{l})$ can be
considered as $|T(\Lambda)|$ modifications of generalized theta
functions. Comparing the expressions (\ref{3.25}) and
(\ref{3.26}), one finds the relation \be \label{3.28}
Z(\lambda,\underline{l})=\frac{\lambda^{r/2}}{\sqrt{|T(\Lambda)|}}
\sum_{\underline{l}'\in T(\Lambda)}\exp\left(2\pi i\lambda l'_I
Q^{IJ}l_J\right)Z\left(\frac{1}{\lambda},\underline{l}'\right) \ee
which can be regarded as action of a variant of the
Montonen--Olive duality transformation \cite{MonOl}, that is, a
($\lambda\rightarrow \frac{1}{\lambda}$)-analogue of the
$S$-transformation $\tau \rightarrow-\frac{1}{\tau}$ of
electric-magnetic duality \cite{Verl, WitS} applied to the
transition amplitude in the BF-model ({\it cf.} section 6 in
\cite{OlAl2}). Thus the transition amplitudes written as
(\ref{3.25}) and (\ref{3.26}), are expressed by means of
$|T(\Lambda)|$ ``theta functions'' depending on the principal
$U(1)$-bundles $P_\partial$ (on the boundary $\partial M$)
classified by Chern classes $c_\partial=l_It^I\in T(\Lambda)$
where $\{t^I\}$ are generators of $H^2(\partial M,\mathbb{Z})$.
Since for our graph cobordisms $T(\Lambda)=H^2(\partial
M,\mathbb{Z})=\oplus_{I=1}^r\mathbb{Z}_{p(I)}$ (pure torsion), all
principal $U(1)$-bundles on $\partial M$ are flat \cite{Zu1}: \be
\label{3.29} \textnormal{Princ}_{_0} (\partial
M)\overset{c_\partial}{\cong}H^2(\partial M,\mathbb{Z})=
\bigoplus_{I=1}^r\mathbb{Z}_{p(I)}. \ee Due to (\ref{2.16}) for
our case, all flat principal $U(1)$-bundles on $\partial M$ (fixed
by the sets $\left\{l_I\in\overline{0,p(I)-1}|I\in\overline{1,r}
\right\}$ also known as {\it rotation numbers} \cite{FintSt,
SavB2}) are extendable to $M$. These extensions are parametrized
by the group of relative bundles $\textnormal{Princ}(M,\partial M)
\overset{c_{\rm rel}}{\cong}H^2(M,\partial M,\mathbb{Z})$ (or
equivalently by the set of integer parts $\left\{m^I\in\mathbb{Z}|
I\in\overline{1,r}\right\}$ of rational ``magnetic'' fluxes
$\Phi^I_{\rm (mag)}$) in the one-to-one manner due to
(\ref{2.17}).

\noindent{\it Observation 3.2.} The partition sums (\ref{3.25})
and (\ref{3.26}) converge if the intersection matrices $Q_{IJ}$
and $Q^{IJ}$ are positive definite. In section \ref{s4} examples
of graph cobordisms satisfying this condition will be given.

\noindent{\it Observation 3.3.} The passage from (\ref{3.25}) to
(\ref{3.26}) for the transition amplitudes corresponds to an
interchange of strong and weak couplings: (\ref{3.25}) involves
the coupling constants matrix $\frac{1}{\lambda}Q_{IJ}$ whose
non-zero elements are $>1$ (strong coupling); this formula is
related to interaction of ``magnetic'' fluxes passing through
homologically non-trivial closed 2-surfaces $\beta^I$, while the
expression (\ref{3.26}) contains the coupling constants matrix
$\lambda Q^{IJ}$ whose non-zero elements are $<1$ (weak coupling),
and it is related to interaction of ``electric'' fluxes $\Phi^{\rm
(el)}_I(\Bar{m},\underline{l}) $ (through homologically
non-trivial closed 2-cycles $\phi_I$) which mimic presence of
quantized ``electric charges'' always being integer according to
(\ref{3.23}). In the same manner, the ``magnetic'' fluxes captured
by homologically non-trivial 2-cycles $\beta^I$ can be interpreted
as effective quantized ``magnetic charges'' possessing rational
values since $\Phi^I_{\rm (mag)}(\Bar{m},\underline{l})\in
\mathbb{Q}$. Note that these effective ``magnetic charges''
pertain to a specific fixed subset of rational numbers with a
finite collection of different denominators composed only by
products of the topological invariants $p'_1,\dots,p'_{2r+1}$ of
the graph cobordism $M$ (see subsection \ref{s2.3}). Thus from the
BF-analogue of the generalized Dirac quantization conditions
(\ref{3.7}) and (\ref{3.8}) it follows that the ``electric
charges'' come to be integer, while the ``magnetic charges'' are
found to be rational.

\noindent{\it Observation 3.4.} It is interesting to note that the
same formulae (\ref{3.23}) and (\ref{3.24}) can be obtained from
fluxes of basic 2-cocycles $b_I\in H^2(M,\partial M,\mathbb
Z)=\Lambda$ and $f^I\in H^2(M, \mathbb Z)=\Lambda^\#$ through a
general 2-cycle $Z_2$ which is possible to represent in the form
\be \label{3.30}Z_2=m^J\phi_J+l_J\beta^J, \ee using the same ideas
as for writing the expression (\ref{3.11}). Then the flux of the
basic field $b_I$ through $Z_2$ reads \be \label{3.31} <b_I,
m^J\phi_J+l_J\beta^J> = m^JQ_{IJ}+l_J\delta^J_I=
\Phi^{\rm(el)}_I(\Bar{m},\underline{l})\in\mathbb{Z} \ee which
coincides with the ``electric'' flux (\ref{3.23}). Thus the matrix
element $Q_{IJ}=<b_I,\phi_J>\in\mathbb{Z}$ may be interpreted as
an elementary ``electric charge'' imitated by the flux of basic
``electric field'' $b_I$ through the basic 2-cycle
$\phi_J\in\Lambda$. Analogously, $\delta_I^J=<b_I,\beta^J>$ can be
understood as an elementary ``electric charge'' simulated by the
flux of basic 2-cocycle $b_I$ through the basic 2-cycle
$\beta^J\in\Lambda^\#$. In the same way the flux of the dual basic
field $f^I\in\Lambda^\#$ through $Z_2$ is \be \label{3.32} <f^I,
m^J\phi_J+l_J\beta^J> = m^J\delta^I_J+l_JQ^{IJ}=
\Phi_{\rm(mag)}^I(\Bar{m},\underline{l})\in\mathbb{Q}, \ee being
the same as the ``magnetic'' flux (\ref{3.24}). This leads to the
interpretation of matrix element
$Q^{IJ}=<f^I,\beta^J>\in\mathbb{Q}$  as an elementary ``magnetic
charge'' imitated by the flux of basic ``magnetic field'' $f^I$
captured by the basic 2-cycle $\beta^J\in\Lambda^\#$, and of
matrix element $\delta^I_J=<f^I,\phi_J>$ as an elementary
``magnetic charge'' imitated by the flux of basic 2-cocycle $f^I$
through the basic 2-cycle $\phi_J\in\Lambda$.

This picture resembles well aged ideas of Wheeler and Misner
\cite{Whee, WheeMis} about ``charges without charges'' when the
field strength lines are captured by topological handles
(wormholes = topological non-trivialities of the spacetime
manifold). It occurs that the spacetime topology has to be
unexpectedly complex when one is trying to reproduce certain
characteristic features of the real universe. In section \ref{s4}
we propose a concrete model exemplifying the possibility of
dealing with the problems of the number of fundamental
interactions in the universe as well as the hierarchy of their
coupling constants on the purely topological level, but using
rather complicated four-manifolds (graph cobordisms).

\subsection{Upper bounds of the scale factor} \label{s3.3}

In our model, the generalized Dirac quantization conditions
(\ref{3.7}) and (\ref{3.8}), together with the exactness of
cohomological sequence (\ref{2.22}), give upper bounds of the
scale factor $\lambda$ introduced in the action (\ref{3.1}). These
bounds are determined in terms of the topological invariants of
the graph cobordism $M$, but they also depend on the Chern classes
$c_\partial=l_It^I$ of the principal $U(1)$-bundles $P_\partial$
which are fixed on the boundary $\partial M$. Note that the
parameter $\lambda$ does determine the scale factor of the
coupling constants matrix as $\lambda Q^{IJ}$ for weak coupling in
(\ref{3.26}) and $\frac{1}{\lambda}Q_{IJ}$ for strong coupling in
(\ref{3.25}) where the rational and integer intersection matrices
$Q^{IJ}$ and $Q_{IJ}$ give the hierarchy of the corresponding
coupling constants. Due to exactness of the cohomological sequence
(\ref{2.22}) and since $H^2(\partial M,\mathbb{Z})$ is a pure
torsion for any class $\frac{F_{\rm cl}}{2\pi}\in H^2(
M,\mathbb{Z} )$ (\ref{3.11}), there exists such a minimal positive
integer $q_0$ that $q_0\frac{F_{\rm cl}}{2\pi}$ is a certain
element $\frac{B_{ \rm cl}}{2\pi}$ of the group $H^2(M,\partial
M,\mathbb{Z})$ \cite{FintSt, FFU}, {\it i.e} \be \label{3.33}
q_0F_{\rm cl}/2\pi= B_{\rm cl}. \ee (It is obvious that for any
positive integer $k$ the class $kq_0F_{\rm cl}/2\pi$ will
certainly belong to the group $H^2(M,\partial M,\mathbb{Z})$.)
Comparing the relation (\ref{3.33}) and the classical constraint
equation (\ref{3.2}), $F=\lambda B$, we find the upper bound of
$\lambda$, namely $\lambda_0=1/q_0$. Moreover, the scale factor
$\lambda$ is quantized in the sense that $\lambda_{k-1}=1/(kq_0)$,
$k\in \mathbb{N}$. The value of $q_0$ can be found from the
expansion of solutions of the equation (\ref{3.33}) with respect
to the bases $\{b_I\}$ and $\{f^I\}$ of the groups $H^2(M,\partial
M,\mathbb{Z})$ and $H^2(M,\mathbb{Z})$, respectively, \be
\label{3.34} \frac{B_{ \rm cl}}{2\pi}=k^Ib_I, ~ ~
k^I\in\mathbb{Z},~ ~ ~ ~ ~ ~ ~ ~ ~ ~ \ee \be \label{3.35}
\frac{F_{ \rm cl}}{2\pi}=m^Ib_I+l_If^I, ~ ~
m^I\in\mathbb{Z},l_I\in\overline{0,p(I)-1} \ee (generalized Dirac
quantization conditions). A substitution of these solutions into
(\ref{3.33}) yields \be \label{3.36} k^Ib_I=
q_0\left(m^Ib_I+l_If^I\right). \ee This condition would be
satisfied if the term $q_0l_If^I$ pertained to the group $H^2(M,
\partial M,\mathbb{Z})$, {\it i.e.} if such a collection
$\left\{s^I|I\in\overline{1,r}\right\}$ of integers could be found
that \be \label{3.37} q_0l_If^I=s^Ib_I. \ee A pairing of the last
equation with the basis elements $f^J$ of the group $H^2(M,
\mathbb{Z})$ yields \be \label{3.38} s^I=q_0Q^{IJ}l_J. \ee Here
the problem consists of rationality of $Q^{IJ}$ while $\{s^I\}$ is
a collection of integers. Note that the cobordisms under
consideration correspond to graphs shown in figure 8 thus having
the only non-zero elements $Q^{II}$ and $Q^{I\,I\pm 1}$. If for
some value of $J$ the rotation number $l_J=0$, the matrix elements
$Q^{JJ}$ and $Q^{J\,J\pm 1}$ do not enter (\ref{3.38}). If $l_J
\neq 0$, such elements give non-zero contribution. These elements
have the common denominator $\Tilde{P}_J=\textnormal{LCM}\,(
p'_{2J-1},p'_{2J},p'_{2J+1})$ where LCM means Least Common
Multiple, while $p'_s$ are the positive integers characterizing
the decorated graph $\Gamma_D$ in figure 8 (see subsections
\ref{s2.1}, \ref{s2.3}). Thus all terms in the right-hand side of
(\ref{3.38}) will be integers, if \be \label{3.39} q_0\equiv q_0(
\underline{l})=\textnormal{LCM}\left(\frac{\Tilde{P}_J}{
\textnormal{GCD}\,(\Tilde{P}_J,l_J)},\textnormal{ over all $J$
such that }l_J\neq 0\right) \ee where GCD means Greatest Common
Divisor, and the notation $q_0(\underline{l})$ takes into account
dependence on the rotation numbers $l_J$. It is worth being
emphasized that the upper bound $\lambda_0(\underline{l})=1/q_0(
\underline{l})$ of the scale factor in our BF-model depends not
only on topological invariants $p'_s$ of the graph cobordism $M$,
but also on the Chern class $c_\partial=l_It^I$ which fixes the
principal bundle $P_\partial$ on the boundary $\partial M$. Note
that $\lambda_0(\underline{l})=1/q_0(\underline{l})$ itself is the
upper bound of the sequence of admissible scale factors
$\lambda_{k-1}(\underline{l})=1/(kq_0( \underline{l})),\,k\in
\mathbb{N}$.

\noindent{\it Observation 3.5} From (\ref{3.39}) one sees that the
quantity $q_0(\underline{l})$ takes its maximum value when all
$l_J\neq 0$ and $\textnormal{GCD}\,(\Tilde{P}_J,l_J)=1$. Then \be
\label{3.40} \Bar{q}_0:=\displaystyle\max_{\underline{l}}
q_0(\underline{l})=\textnormal{LCM}\,(\tilde{P}_J,J=\overline{1,r})
=\textnormal{LCM}\,(p'_s,s=\overline{1,2r+1}). \ee The quantity
$q_0(\underline{l})$ takes its minimum value when all $l_J=0$. It
is obvious that in this case $\underline{q}_0:=\displaystyle\min_{
\underline{l}} q_0(\underline{l})=1$. Thus the upper bounds
$\lambda_0(\underline{l})$ of the scale factor in our BF-model
take discrete values in the interval \be \label{3.41}
\frac{1}{\Bar{q}_0}\leq\lambda_0(\underline{l})\leq 1. \ee In
section \ref{s4} we shall build a cosmological model in which the
present-stage universe is characterized by an integer $\Bar{q}_0$
having the order of magnitude $3.28\cdot 10^{177}$.

\section{The family of graph cobordisms as a
sequence of cosmological models} \label{s4}
\setcounter{equation}{0}

In this section we construct a collection of graph cobordisms
interpretable as a sequence of topological changes finally
resulting in the state of universe which we identify as its
contemporary stage by the number of fundamental interactions and
the hierarchy of their coupling constants. We proposed a
similar type of model in recent papers \cite{EM, Stan}. The
construction we realize now differs by an additional condition on
the four-dimensional topological space playing the r\^ole of the
spacetime manifold: its intersection matrix is demanded to be
positive defined (see observation 2.2). This guarantees
convergence of partition sums (\ref{3.25}) and (\ref{3.26}) in the
topological gauge theories built on the graph cobordisms (see
section \ref{s3}).

\subsection{The basic family of Seifert fibred homology spheres}
\label{s4.1}

The basic structure elements of graph cobordisms used in this
paper are simple graph four-manifolds with Seifert fibred
Brieskorn homology (Bh-) spheres $\Sigma(a_1,a_2,a_3)$ as
boundaries (see subsection \ref{s2.1}). (Compact locally
homogeneous universes with spatial sections homeomorphic to
Seifert fibrations were considered at length in \cite{KTH1, KTH2,
Fag}.) We use only a specific bi-parametric family of Bh-spheres
which is defined as follows: First, we introduce the {\it primary
sequence} of Bh-spheres (see \cite{EM} for more details). Let
$p_i$ be the $i$th prime number in the set of positive integers
$\mathbb{N}$, {\it e.g.} $p_1=2, ~ p_2=3,\dots, p_9=23,\dots$.
Then the primary sequence is defined as \be \label{4.1} \left\{
\Sigma(p_{2n},p_{2n+1},q_{2n-1})|n\in \mathbb{Z}^+\right\} \ee
where $q_i:=p_1\cdots p_i$, $\mathbb{Z}^+$ is the set of
non-negative integers. The first terms in this sequence with $n>0$
(which we really use) are $\Sigma(2,3,5)$ (the Poincar\'e homology
sphere), $\Sigma(7,11,30)$, $\Sigma(13,17, 2310)$, and
$\Sigma(19,23,510510)$. We also include in this sequence as its
first term ($n=0$) the usual three-dimensional sphere $S^3$
(Sf-sphere) with Seifert fibration determined by the mapping
$h_{pq}:S^3\rightarrow S^2$, in its turn defined as $h_{pq}(z_1,
z_2)=z^p_1/z^q_2$ \cite{Scott}. Recall that
$S^3=\{(z_1,z_2)||z_1|^2+|z_2|^2=1\}$ and
$z^p_1/z^q_2\in\mathbb{C}\cup\{\infty\}\cong S^2$. In this paper
we consider the case $p=1$, $q=2$ and denote this Sf-sphere as
$\Sigma(1,2,1)$, {\it i.e.} $p_0=q_{-1}=1$ in (\ref{4.1}). In this
notation we use two additional units which correspond to two
arbitrary regular fibers. This will enable us to operate with
$\Sigma(1,2,1)$ in the same manner as with other members of the
sequence (\ref{4.1}).

Second, we define $k^\pm$-operations for each of Bh-spheres
$\Sigma(a_1,a_2,a_3)$. To start with, we renumber Seifert's
invariants so that $a_1<a_2<a_3$; this is always possible since
$a_1$, $a_2$ and $a_3$ are pairwise coprime (in the case of the
Sf-sphere, we take the order $\Sigma(1,1,2)$). The result of
$k^\pm$-operation acting on $\Sigma(a_1,a_2,a_3)$ is another
Bh-sphere \be \label{4.2} \Sigma^\pm_{k_1}(a^{(1)}_1,a^{(1)}_2,
a^{(1)}_3)=\Sigma(a_1,a_2a_3,k_1a\pm1), \ee {\it i.e.} it is the
Bh-sphere with Seifert invariants \be \label{4.3} a^{(1)}_1=a_1,
a^{(1)}_2=a_2a_3, a^{(1)}_3=k_1a\pm1 \ee where $a=a_1a_2a_3$, $k_1
\in\mathbb{N}$. The upper index in the parentheses means a single
application of the $k^\pm$-operation. A repeated application of
this operation yields still another Bh-sphere \be \label{4.4}
\Sigma^\pm_{k_1k_2}(a^{(2)}_1,a^{(2)}_2, a^{(2)}_3)=
\Sigma(a_1,a_2a_3(k_1a\pm1),k_2a(k_1a\pm1)\pm1) \ee where $k_2\in
\mathbb{N}$; in general, $k_2\neq k_1$. The $l$-fold application
of the $k^\pm$-operation again gives an Bh-sphere, $\Sigma^\pm_{
k_1\dots k_l}(a^{(l)}_1,a^{(l)}_2, a^{(l)}_3)$ whose invariants
are found by induction from the invariants $a^{(l-1)}_1,a^{
(l-1)}_2,a^{(l-1)}_3$, with arbitrary $k_l\in\mathbb{N}$. Note
that the least Seifert invariant does not change under
$k^\pm$-operations ($a^{(l)}_1=a_1$ for any $l=1,2,\dots$) while
the two other Seifert invariants depend both on the order
(multiplicity) of the $k^\pm$-operation fulfilment and on which
($k^+$ or $k^-$)-operation is applied. A hint of such an operation
can be found in Saveliev's paper \cite{SavRuss}.

In \cite{EM} we defined only the $k^+$-operation in the special
case $k_1=k_2=\cdots=k_l=1$ and named it (not quite aptly)
``derivative of Bh-sphere''. In our new terminology this is the
$1^+$-operation; its $l$-fold application gives the Bh-sphere
denoted in \cite{EM} as $\Sigma(a^{(l)}_1,a^{(l)}_2, a^{(l)}_3)$.
In the same paper we showed that the application of this operation
to the primary sequence (\ref{4.1}) yields a bi-parametric family
of Bh-spheres whose Euler numbers reproduce fairly well the
experimental hierarchy of dimensionless low-energy coupling (DLEC)
constants of the fundamental interactions in the real universe.
For the reader's convenience we concisely reiterate here some
results obtained in \cite{EM}.

This bi-parametric family of Bh-spheres is \be \label{4.5}
\left\{\Sigma(a^{(l)}_{1n},a^{(l)}_{2n},a^{(l)}_{3n})= \Sigma(p^{
(l)}_{2n},p^{(l)}_{2n+1},q^{(l)}_{2n-1})|n,l\in
\mathbb{Z}^+\right\}. \ee (Note that the $k^\pm$-operation
involves a renumbering of Seifert's invariants such that the
inequalities $ a^{(l)}_{1n}< a^{(l)}_{2n}<a^{(l)}_{3n}$ become
valid. Thus the collections of Seifert's invariants
$\left\{a^{(l)}_{1n},a^{(l)}_{2n},a^{(l)}_{3n}\right\}$ and
$\left\{p^{ (l)}_{2n},p^{(l)}_{2n+1},q^{(l)}_{2n-1}\right\}$ are
equivalent up to ordering.) In \cite{EM} it was shown that to
reproduce the hierarchy of the DLEC constants of the known five
fundamental interactions (including the cosmological one) it is
sufficient to restrict values of parameters as $n,l\in\overline{
0,4}$. With this restriction, the Euler numbers of the Bh-spheres
family are given in table \ref{tab:1} (the revised table 3 of
\cite{EM}).

\begin{table*}{\scriptsize
\caption{\label{tab:1}Euler number of $(n,t)$-family of Sf- and
Bh-spheres.}
~~~~~\\
\begin{tabular}{|c|l|l|l|l|l|l|l|l|l|}
\hline\hline
$\hspace*{-.4cm}_{n}$\hspace*{-.2cm}$\diagdown$\hspace*{-.1cm}$^{t}
$\hspace*{-.4cm} & $-4$ & $-3$ & $-2$ & $-1$ & 0 & 1 & 2 & 3 &
4\\\hline\hline 0 &  &  &  &  &\hspace*{-.2cm}
$\mathbf{5.0\hspace*{-.1cm}\times 10^{-1}}$\hspace*{-.2cm}
&\hspace*{-.2cm}
$1.7\hspace*{-.1cm}\times\hspace*{-.1cm}10^{-1}$\hspace*{-.2cm}
&\hspace*{-.2cm}
$2.3\hspace*{-.1cm}\times\hspace*{-.1cm}10^{-2}$\hspace*{-.2cm}
&\hspace*{-.2cm}
$5.5\hspace*{-.05cm}\times\hspace*{-.1cm}10^{-4}$\hspace*{-.2cm}
&\hspace*{-.2cm} $3.1\hspace*{-.1cm}\times\hspace*{-.1cm}10^{-7}
$\hspace*{-.2cm}\\\hline 1 &  &  &  &\hspace*{-.2cm}
$3.3\hspace*{-.1cm}\times\hspace*{-.1cm}10^{-2}$ \hspace*{-.2cm}&
\hspace*{-.1cm}$\mathbf{1.1\hspace*{-.1cm}\times\hspace*{-.1cm}10^{-3}}
$\hspace*{-.2cm} &\hspace*{-.2cm}
$1.2\hspace*{-.1cm}\times\hspace*{-.1cm}10^{-6}$
\hspace*{-.2cm}&\hspace*{-.2cm}
$1.3\hspace*{-.1cm}\times\hspace*{-.1cm}10^{-12}$
\hspace*{-.2cm}&\hspace*{-.2cm}
$1.8\hspace*{-.1cm}\times\hspace*{-.1cm}10^{-24}$\hspace*{-.2cm}
&\hspace*{-.2cm} \\\hline 2 &  &  &\hspace*{-.2cm}
$4.3\hspace*{-.1cm}\times\hspace*{-.1cm}10^{-4}$\hspace*{-.2cm}
&\hspace*{-.2cm}
$1.9\hspace*{-.1cm}\times\hspace*{-.1cm}10^{-7}$\hspace*{-.2cm}
&\hspace*{-.2cm} $\mathbf{3.5
\hspace*{-.1cm}\times\hspace*{-.1cm}10^{-14}}$\hspace*{-.2cm} &
\hspace*{-.1cm}$1.2\hspace*{-.1cm}\times\hspace*{-.1cm}10^{-27}
$\hspace*{-.2cm} &
\hspace*{-.1cm}$1.5\hspace*{-.1cm}\times\hspace*{-.1cm}10^{-54}
$\hspace*{-.2cm} &  & \\\hline 3 &  &
\hspace*{-.1cm}$2.0\hspace*{-.1cm}\times\hspace*{-.1cm}10^{-6}
$\hspace*{-.2cm} &\hspace*{-.2cm}
$3.8\hspace*{-.1cm}\times\hspace*{-.1cm}10^{-12}$
\hspace*{-.2cm}&\hspace*{-.2cm} $1.5
\hspace*{-.1cm}\times\hspace*{-.1cm}10^{-23}$\hspace*{-.2cm}
&\hspace*{-.2cm}
$\mathbf{2.2\hspace*{-.1cm}\times\hspace*{-.1cm}10^{-46}}$\hspace*{-.2cm}
&\hspace*{-.2cm}
$4.7\hspace*{-.1cm}\times\hspace*{-.1cm}10^{-92}$\hspace*{-.2cm} &
&  & \\\hline 4 &
\hspace*{-.15cm}$4.5\hspace*{-.1cm}\times\hspace*{-.1cm}10^{-9}
$\hspace*{-.2cm} &\hspace*{-.2cm}
$2.0\hspace*{-.1cm}\times\hspace*{-.1cm}10^{-17}$\hspace*{-.2cm}&
\hspace*{-.1cm}$4.0
\hspace*{-.1cm}\times\hspace*{-.1cm}10^{-34}$\hspace*{-.2cm}
&\hspace*{-.2cm} $1.6\hspace*{-.1cm}\times\hspace*{-.1cm}10^{-67}$
\hspace*{-.2cm}&\hspace*{-.2cm}
$\mathbf{2.7\hspace*{-.1cm}\times\hspace*{-.1cm}10^{-134}}$
\hspace*{-.3cm}& &  &  & \\\hline
\end{tabular}}
\end{table*}

To make the comparison with the experimental hierarchy of DLEC
constants (see table \ref{tab:2}) easier, we introduced instead of
$l$ a new parameter $t:=l-n$ which plays the r\^ole of ``discrete
cosmological time''.

\begin{table}[h]
\caption{ \label{tab:2} Euler numbers {\it vs.} experimental DLEC
constants {\it vs.}
\newline
diagonal elements of intersection matrix $Q^+(0)$ (see subsection
\ref{s4.3}).}
\begin{center}
\begin{tabular}[c]{|l|l|l|l|l|l|}\hline\hline
$n$ & {$e\left(  \Sigma^{(n)}_{\phantom{|}n}\right) $} &
Interaction & $\alpha_{{\rm exper}}$ & $Q^{+II}(0)$&$I$\\
\hline $0$ & $0.5$ & strong & $1$ & $9.69\times10^{-1}$ & $1$\\
\hline $1$ & $1.07\times10^{-3}$ & electromagnetic &
$7.20\times10^{-3}$ & $7.21\times10^{-3}$ & $2$\\
\hline $2$ & $3.51\times10^{-14}$ & weak &
$3.04\times10^{-12}$&$1.76\times10^{-12}$&$3$\\
\hline $3$ & $2.17\times10^{-46}$ &
gravitational & $2.73\times10^{-46}$&$3.68\times10^{-44}$&$4$\\
\hline $4$ & $2.70\times10^{-134}$ & cosmological & $<10^{-120}$
&$2.66\times10^{-134}$&$5$\\
\hline \end{tabular}
\end{center}
~~~ \\
 {\footnotesize
{\bf Notes:} {\bf 1.} The dimensionless strong interaction
constant is $\alpha_{\mathrm{st}}= G/\hbar c$, $G$ characterizes
the strength of the coupling of the meson field to the nucleon.
{\bf 2.} The fine structure (electromagnetic) constant is
$\alpha_{\mathrm{em}}=e^2/\hbar c$. {\bf 3.} The dimensionless
weak interaction constant is
$\alpha_{\mathrm{weak}}=(G_{\mathrm{F}}/\hbar c)(m_{\mathrm
e}c/\hbar)^2$, $G_{\mathrm{F}}$ being the Fermi constant
($m_{\mathrm e}$ is mass of electron). {\bf 4.} The dimensionless
gravitational coupling constant is
$\alpha_{\mathrm{gr}}=G_Nm_{\mathrm e}^2/\hbar c$,
$G_{\mathrm{N}}$ being the Newtonian gravitational constant. {\bf
5.} The cosmological constant $\Lambda$ multiplied by the squared
Planckian length is $\alpha_{\mathrm{cosm}}=\Lambda
G_{\mathrm{N}}\hbar /c^3$.  The mentioned dimensionless constants
(except the cosmological one) are also known as Dyson numbers.}
\end{table}
Just at $t=0$ ($l=n$) the experimental hierarchy of DLEC constants
is reproduced properly. This enables us to consider the ensemble
of Bh-spheres (\ref{4.5}) at $l=n$ \be \label{4.6} E_0=
\left\{\Sigma(a^{(n)}_{1n},a^{(n)}_{2n},a^{(n)}_{3n})= \Sigma(p^{
(n)}_{2n},p^{(n)}_{2n+1},q^{(n)}_{2n-1})|n\in\overline{0,4}\right\}
\ee as the basis elements used in constructing the spatial section
$\Sigma_0$ of the contemporary universe by means of the splicing
operation. The key factor in this (at first glance, exotic)
hypothesis is the fact that the diagonal elements (and
eigenvalues) of the rational intersection matrix for the
corresponding graph cobordism $M_0$ (that is, its $\partial M_0=
\Sigma_0\bigsqcup_{s=1}^N\left(-L(|p'_s|,q'_s)\right)$) show the
same hierarchy as the Euler numbers, thus reproducing the DLEC
constants' hierarchy (see two last columns in table 2). Then it is
natural to suppose that at $t\in \overline{-4,-1}$ the ensembles
\be \label{4.7} E_t
=\left\{\Sigma(a^{(n+t)}_{1n},a^{(n+t)}_{2n},a^{(n+t)}_{3n})=
\Sigma(p^{ (n+t)}_{2n},p^{(n+t)}_{2n+1},q^{(n+t)}_{2n-1})|n\in
\overline{-t,4}\right\} \ee of Bh-spheres forming basic elements
for gluing (by splicing) spatial sections of the universe on
earlier stages characterized, in particular, by a diminishing of
the number of fundamental interactions from five at $t=0$ to one
at $t=-4$.

\begin{figure}[h]
\begin{center}
\setlength{\unitlength}{1pt}
\begin{picture}(150,50)

\put(225,-.5){\makebox(.5,.5){$\bullet$}}
\put(225,-50.5){\makebox(.5,.5){$\bullet$}}
\put(225,-100.5){\makebox(.5,.5){$\bullet$}}
\put(225,-150.5){\makebox(.5,.5){$\bullet$}}
\put(225,-200.5){\makebox(.5,.5){$\bullet$}}
\put(200,25){\makebox(.5,.5){$\bullet$}}
\put(200,-225){\makebox(.5,.5){$\bullet$}}

\put(211,3){\makebox(3,3){\scriptsize $a_{10}$}}
\put(191,10){\makebox(3,3){\scriptsize $a_{20}$}}
\put(191,-13){\makebox(3,3){\scriptsize $a_{30}$}}

\put(211,-43){\makebox(3,3){\scriptsize $a^{(1)}_{11}$}}
\put(191,-40){\makebox(3,3){\scriptsize $a^{(1)}_{31}$}}
\put(191,-63){\makebox(3,3){\scriptsize $a^{(1)}_{21}$}}

\put(211,-93){\makebox(3,3){\scriptsize $a^{(2)}_{12}$}}
\put(191,-90){\makebox(3,3){\scriptsize $a^{(2)}_{32}$}}
\put(191,-113){\makebox(3,3){\scriptsize $a^{(2)}_{22}$}}

\put(211,-143){\makebox(3,3){\scriptsize $a^{(3)}_{13}$}}
\put(191,-140){\makebox(3,3){\scriptsize $a^{(3)}_{33}$}}
\put(191,-163){\makebox(3,3){\scriptsize $a^{(3)}_{23}$}}

\put(211,-193){\makebox(3,3){\scriptsize $a^{(4)}_{14}$}}
\put(191,-190){\makebox(3,3){\scriptsize $a^{(4)}_{34}$}}
\put(191,-213){\makebox(3,3){\scriptsize $a^{(4)}_{24}$}}

\put(211,-63){\makebox(3,3){\scriptsize $k^{(1)}_{1}$}}
\put(211,-113){\makebox(3,3){\scriptsize $k^{(2)}_{2}$}}
\put(211,-163){\makebox(3,3){\scriptsize $k^{(3)}_{3}$}}
\put(211,-213){\makebox(3,3){\scriptsize $k^{(4)}_{4}$}}

\put(205,0){\line(1,0){20}}
\put(200,5){\line(0,1){20}}
\put(200,-5){\line(0,-1){20}}
\put(200,0){\circle{10}}
\put(199.5,-.50){\makebox(.5,.5){$-$}}

\put(205,-50){\line(1,0){20}}
\put(200,-45){\line(0,1){20}}
\put(200,-55){\line(0,-1){20}}
\put(200,-50){\circle{10}}
\put(199.5,-50.50){\makebox(.5,.5){$-$}}

\put(205,-100){\line(1,0){20}}
\put(200,-95){\line(0,1){20}}
\put(200,-105){\line(0,-1){20}}
\put(200,-100){\circle{10}}
\put(199.5,-100.50){\makebox(.5,.5){$-$}}

\put(205,-150){\line(1,0){20}}
\put(200,-145){\line(0,1){20}}
\put(200,-155){\line(0,-1){20}}
\put(200,-150){\circle{10}}
\put(199.5,-150.50){\makebox(.5,.5){$-$}}

\put(205,-200){\line(1,0){20}}
\put(200,-195){\line(0,1){20}}
\put(200,-205){\line(0,-1){20}}
\put(200,-200){\circle{10}}
\put(199.5,-200.50){\makebox(.5,.5){$-$}}


\put(165,-50.5){\makebox(.5,.5){$\bullet$}}
\put(165,-100.5){\makebox(.5,.5){$\bullet$}}
\put(165,-150.5){\makebox(.5,.5){$\bullet$}}
\put(165,-200.5){\makebox(.5,.5){$\bullet$}}
\put(140,-25){\makebox(.5,.5){$\bullet$}}
\put(140,-225){\makebox(.5,.5){$\bullet$}}

\put(151,-45){\makebox(3,3){\scriptsize $a_{11}$}}
\put(131,-40){\makebox(3,3){\scriptsize $a_{21}$}}
\put(131,-63){\makebox(3,3){\scriptsize $a_{31}$}}

\put(151,-93){\makebox(3,3){\scriptsize $a^{(1)}_{12}$}}
\put(131,-90){\makebox(3,3){\scriptsize $a^{(1)}_{32}$}}
\put(131,-113){\makebox(3,3){\scriptsize $a^{(1)}_{22}$}}

\put(151,-143){\makebox(3,3){\scriptsize $a^{(2)}_{13}$}}
\put(131,-140){\makebox(3,3){\scriptsize $a^{(2)}_{33}$}}
\put(131,-163){\makebox(3,3){\scriptsize $a^{(2)}_{23}$}}

\put(151,-193){\makebox(3,3){\scriptsize $a^{(3)}_{14}$}}
\put(131,-190){\makebox(3,3){\scriptsize $a^{(3)}_{34}$}}
\put(131,-213){\makebox(3,3){\scriptsize $a^{(3)}_{24}$}}

\put(151,-113){\makebox(3,3){\scriptsize $k^{(1)}_{2}$}}
\put(151,-163){\makebox(3,3){\scriptsize $k^{(2)}_{3}$}}
\put(151,-213){\makebox(3,3){\scriptsize $k^{(3)}_{4}$}}

\put(145,-50){\line(1,0){20}}
\put(140,-45){\line(0,1){20}}
\put(140,-55){\line(0,-1){20}}
\put(140,-50){\circle{10}}
\put(139.5,-50.50){\makebox(.5,.5){$-$}}

\put(145,-100){\line(1,0){20}}
\put(140,-95){\line(0,1){20}}
\put(140,-105){\line(0,-1){20}}
\put(140,-100){\circle{10}}
\put(139.5,-100.50){\makebox(.5,.5){$-$}}

\put(145,-150){\line(1,0){20}}
\put(140,-145){\line(0,1){20}}
\put(140,-155){\line(0,-1){20}}
\put(140,-150){\circle{10}}
\put(139.5,-150.50){\makebox(.5,.5){$-$}}

\put(145,-200){\line(1,0){20}}
\put(140,-195){\line(0,1){20}}
\put(140,-205){\line(0,-1){20}}
\put(140,-200){\circle{10}}
\put(139.5,-200.50){\makebox(.5,.5){$-$}}


\put(105,-100.5){\makebox(.5,.5){$\bullet$}}
\put(105,-150.5){\makebox(.5,.5){$\bullet$}}
\put(105,-200.5){\makebox(.5,.5){$\bullet$}}
\put(80,-75){\makebox(.5,.5){$\bullet$}}
\put(80,-225){\makebox(.5,.5){$\bullet$}}

\put(91,-96){\makebox(3,3){\scriptsize $a_{12}$}}
\put(71,-90){\makebox(3,3){\scriptsize $a_{22}$}}
\put(71,-113){\makebox(3,3){\scriptsize $a_{32}$}}

\put(91,-143){\makebox(3,3){\scriptsize $a^{(1)}_{13}$}}
\put(71,-140){\makebox(3,3){\scriptsize $a^{(1)}_{33}$}}
\put(71,-163){\makebox(3,3){\scriptsize $a^{(1)}_{23}$}}

\put(91,-193){\makebox(3,3){\scriptsize $a^{(2)}_{14}$}}
\put(71,-190){\makebox(3,3){\scriptsize $a^{(2)}_{34}$}}
\put(71,-213){\makebox(3,3){\scriptsize $a^{(2)}_{24}$}}

\put(91,-163){\makebox(3,3){\scriptsize $k^{(1)}_{3}$}}
\put(91,-213){\makebox(3,3){\scriptsize $k^{(2)}_{4}$}}

\put(85,-100){\line(1,0){20}}
\put(80,-95){\line(0,1){20}}
\put(80,-105){\line(0,-1){20}}
\put(80,-100){\circle{10}}
\put(79.5,-100.50){\makebox(.5,.5){$-$}}

\put(85,-150){\line(1,0){20}}
\put(80,-145){\line(0,1){20}}
\put(80,-155){\line(0,-1){20}}
\put(80,-150){\circle{10}}
\put(79.5,-150.50){\makebox(.5,.5){$-$}}

\put(85,-200){\line(1,0){20}}
\put(80,-195){\line(0,1){20}}
\put(80,-205){\line(0,-1){20}}
\put(80,-200){\circle{10}}
\put(79.5,-200.50){\makebox(.5,.5){$-$}}


\put(45,-150.5){\makebox(.5,.5){$\bullet$}}
\put(45,-200.5){\makebox(.5,.5){$\bullet$}}
\put(20,-125){\makebox(.5,.5){$\bullet$}}
\put(20,-225){\makebox(.5,.5){$\bullet$}}

\put(31,-146){\makebox(3,3){\scriptsize $a_{13}$}}
\put(11,-140){\makebox(3,3){\scriptsize $a_{23}$}}
\put(11,-163){\makebox(3,3){\scriptsize $a_{33}$}}

\put(31,-193){\makebox(3,3){\scriptsize $a^{(1)}_{14}$}}
\put(11,-190){\makebox(3,3){\scriptsize $a^{(1)}_{34}$}}
\put(11,-213){\makebox(3,3){\scriptsize $a^{(1)}_{24}$}}

\put(31,-213){\makebox(3,3){\scriptsize $k^{(1)}_{4}$}}

\put(25,-150){\line(1,0){20}}
\put(20,-145){\line(0,1){20}}
\put(20,-155){\line(0,-1){20}}
\put(20,-150){\circle{10}}
\put(19.5,-150.50){\makebox(.5,.5){$-$}}

\put(25,-200){\line(1,0){20}}
\put(20,-195){\line(0,1){20}}
\put(20,-205){\line(0,-1){20}}
\put(20,-200){\circle{10}}
\put(19.5,-200.50){\makebox(.5,.5){$-$}}


\put(-15,-200.5){\makebox(.5,.5){$\bullet$}}
\put(-40,-175){\makebox(.5,.5){$\bullet$}}
\put(-40,-225){\makebox(.5,.5){$\bullet$}}

\put(-29,-196){\makebox(3,3){\scriptsize $a_{14}$}}
\put(-49,-190){\makebox(3,3){\scriptsize $a_{34}$}}
\put(-49,-213){\makebox(3,3){\scriptsize $a_{24}$}}

\put(-35,-200){\line(1,0){20}}
\put(-40,-195){\line(0,1){20}}
\put(-40,-205){\line(0,-1){20}}
\put(-40,-200){\circle{10}}
\put(-40.5,-200.50){\makebox(.5,.5){$-$}}


\put(200,40){\makebox(3,3){\scriptsize $t=0$}}
\put(140,40){\makebox(3,3){\scriptsize $t=-1$}}
\put(80,40){\makebox(3,3){\scriptsize $t=-2$}}
\put(20,40){\makebox(3,3){\scriptsize $t=-3$}}
\put(-40,40){\makebox(3,3){\scriptsize  $t=-4$}}

\put(-90,0){\makebox(3,3){\scriptsize $n=0$}}
\put(-90,-50){\makebox(3,3){\scriptsize $n=1$}}
\put(-90,-100){\makebox(3,3){\scriptsize $n=2$}}
\put(-90,-150){\makebox(3,3){\scriptsize $n=3$}}
\put(-90,-200){\makebox(3,3){\scriptsize $n=4$}}

\end{picture}\\
\vspace*{8.5cm} \caption{The splice
diagram of different states of the universe.}
\end{center}
\end{figure}
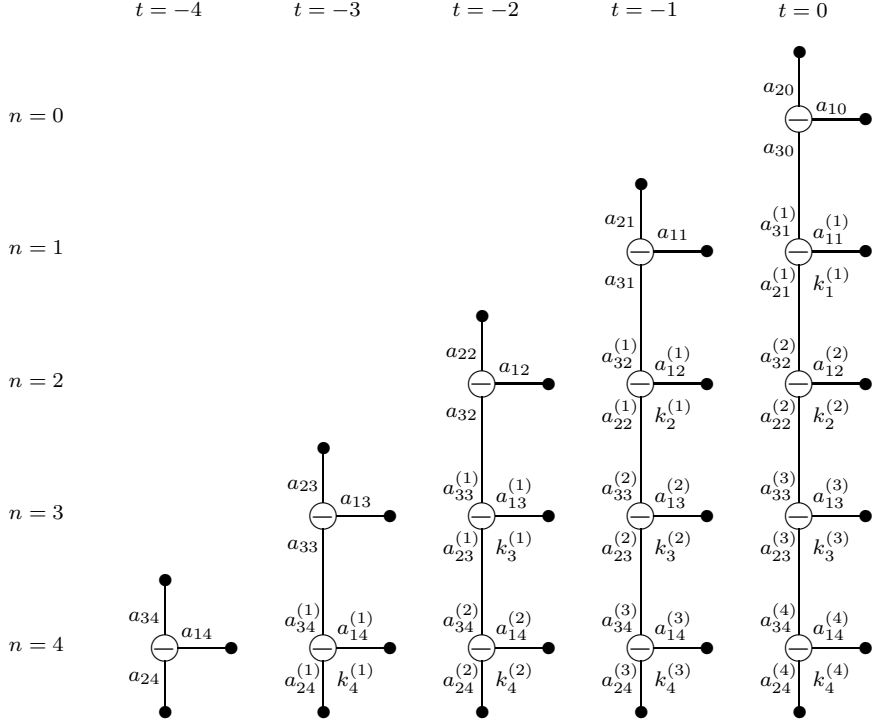

\noindent{\it Observation 4.1.} In this paper we shall not
consider splice diagrams occurring for $t>0$. Just note that in
accordance with table \ref{tab:1} the number of interactions
should diminish from five to one with the increase of the
parameter $t$ from 0 to 4, but the intersection matrices (related
to the coupling constants ones) are different in ascending and
descending stages.

\noindent{\it Observation 4.2.}It is worth being observed that in
our scheme the five (low energy) interactions are related to the
first nine prime numbers as (1,2), (3,5), (7,11), (13,17),
(19,23). To obtain any new interaction, one has to attach a new
pair of prime numbers to the preceding set. For example, taking
the next pair (29,31), we come with the same algorithm to a new
coupling constant of the order of magnitude $\alpha_6\approx
10^{-361}$. Thus our model answers the question we did not even
put: How many fundamental interactions may really exist in the
universe? Our model predicts an infinite number of interactions
due to the infinite succession of prime numbers. We simply cannot
detect too weak interactions beginning with $\alpha_6$ since all
subsequent are even weaker: $\alpha_7\approx 10^{-916}$, {\it
etc.}  \cite{MG11}.

\begin{figure}[h]
\begin{center}
\setlength{\unitlength}{1pt}
\begin{picture}(150,50)


\put(85,4.5){\makebox(.5,.5){$\bullet$}}
\put(85,-45.5){\makebox(.5,.5){$\bullet$}}

\put(71,9){\makebox(3,3){\scriptsize $a_{1n}$}}
\put(51,15){\makebox(3,3){\scriptsize $a_{3n}$}}
\put(51,-8){\makebox(3,3){\scriptsize $a_{2n}$}}

\put(71,-38){\makebox(3,3){\scriptsize $a_{1m}$}}
\put(51,-35){\makebox(3,3){\scriptsize $a_{3m}$}}
\put(51,-58){\makebox(3,3){\scriptsize $a_{2m}$}}

\put(65,5){\line(1,0){20}}
\put(60,10){\line(0,1){20}}
\put(60,0){\line(0,-1){20}}
\put(60,5){\circle{10}}
\put(59.5,4.5){\makebox(.5,.5){$-$}}

\put(65,-45){\line(1,0){20}}
\put(60,-40){\line(0,1){20}}
\put(60,-50){\line(0,-1){20}}
\put(60,-45){\circle{10}}
\put(59.5,-45.5){\makebox(.5,.5){$-$}}

\multiput(58.5,35)(0,3){3}{.}
\multiput(58.5,-82)(0,3){3}{.}

\end{picture}\\
\vspace*{3cm}
\caption{A portion of a splice diagram.}
\end{center}
\end{figure}
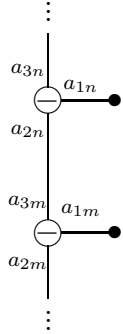

\subsection{The construction of graph cobordisms} \label{s4.2}

The splice diagrams corresponding to states of the universe at the
cosmological time $t\in\overline{-4,0}$ are shown in figure 9
where we consider them as subdiagrams being parts of a disjoint
total splice diagram. To any of these subdiagrams one associates
(in accordance with the well-known algorithm, \cite{EisNeum,
SavArt, NeumWahl}) the graph cobordism $M_D$ (constructed in
Observation 2.2, subsection \ref{s2.1}) whose intersection form is
positive definite iff for each edge joining two nodes the edge
determinant is positive. The {\it edge determinant} $\det(e_{mn})$
of an edge joining two nodes is the product of the two weights on
the edge minus the product of the weights adjacent to the edge
\cite{NeumWahl}. In our case this criterion means that in any
portion of splice diagram shown in figure 10, {\it i.e.} for any
edge $e_{mn}$ one has \be \label{4.8} \det
(e_{mn})=a_{2n}a_{3m}-a_{1n}a_{3n}a_{1m}a_{2m}>0, \ee {\it cf.}
subsection \ref{s2.1} where a more general form of the edge
determinant is given, $\det(e_{mn})=-p$, $p$ being defined in
(\ref{2.3}). Taking certain collections of integers $k_1,\dots,
k_4$ (which participate in $k^\pm$-operations), one gets positive
definite intersection forms for all splice diagrams shown in
figure 9. Note that when $k_1=k_2=k_3=k_4=1$, all intersection
matrices have indefinite signature, thus convergence of transition
amplitudes (\ref{3.25}) and (\ref{3.26}) for corresponding graph
cobordisms is not ensured. Below we confine ourselves to the
$k^+$-operation (for the $k^-$-operation we shall only give the
final result in the Appendix).

The total disconnected diagram in figure 9 consists of five
connected splice subdiagrams. Naturally, there exists an ambiguity
in the splice operation (related to this type of diagrams). The
diagram in figure 9 contains fifteen nodes, each of them having
three adjacent edges. Thus one can glue $3^{15}$ different graph
cobordisms. Moreover, there is an infinite set of integers $k_i$
which guarantee positive definiteness of the intersection matrices
of respective cobordisms. It is however possible to fix a unique
gluing procedure imposing a minimality condition on the
coefficients $k_i$ at each level of realization of the
$k^+$-operation. In particular, this condition immediately yields
a conclusion that the vertices (leaves) corresponding to minimal
Seifert invariants ($a^{(l)}_1$) remain free (not subjected to
slicing). Applying the $k^+$-operation to all Bh-spheres of the
primary sequence \be \label{4.9} \left\{\Sigma(a_{1n},a_{2n},
a_{3n})|n\in\overline{1,4}\right\}, \ee we find the minimal
$k^{(1)}_n$ for which the conditions $a_{3n}a^{(1)}_{3,n+1}-a_{1n}
a_{2n}a^{(1)}_{1,n+1}a^{(1)}_{2,n+1}\linebreak >0$,\,
$n\in\overline{0,3}$ are satisfied. Thus we unambiguously fixed
the collection of the first-level Bh-spheres, {\it i.e.} those
with the parameter $l=1$: \be \label{4.10}
\left\{\Sigma_{k^{(1)}_n}
(a^{(1)}_{1n},a^{(1)}_{2n},a^{(1)}_{3n})|n\in\overline{1,4}
\right\}. \ee Now we execute the first splicing procedure (along
the upper vertical edges between nodes in figure 9): \be
\label{4.11} \Sigma(a_{1n},a_{2n}, a_{3n})\frac{~}{S ~ S'}
\Sigma_{k^{(1)}_{n+1}} (a^{(1)}_{1,n+1},a^{(1)}_{2,n+1},
a^{(1)}_{3,n+1}) \ee where $n\in\overline{0,3}$, $S=S_{a_{3n}}$,
$S'=S_{a_{3,n+1}}$. The same algorithm is applied to determine the
collection of the second-level Bh-spheres, and further by
induction the $l$th-level Bh-spheres: \be \label{4.12}
\left\{\Sigma_{k^{(1)}_n\dots k^{(l)}_n}
(a^{(l)}_{1n},a^{(l)}_{2n},a^{(l)}_{3n})|n\in\overline{l,4}
\right\}. \ee In each step, there is executed the splice operation
according to the diagram in figure 9.

Consequently, we obtain the five connected subdiagrams
$\Delta^+(t)$ where superscript $~^+$ corresponds to the use of
$k^+$-operation with minimization of parameters $k^{(l)}_n$ at
each step. According to the procedure described in the subsection
\ref{s2.1}, the corresponding decorated plumbed graphs
$\Gamma^+_D(t)$ are constructed. These decorated graphs codify the
definite graph cobordisms $M^+_D(t)$, $t\in\overline{-4,0}$, which
are interpreted as the spacetime manifolds corresponding to
different values of cosmological time parameter $t$. The
boundaries of these cobordosms are represented as follows: \be
\label{4.13} \partial
M^+_D(t)=\left(-\bigsqcup_{s=1}^{N(t)}L(|p'_s(t)|,q'_s(t))
\right)\bigsqcup \Sigma^+(t). \ee (see the expression
(\ref{2.8})). It is worth being underlined that both $\mathbb
Z$-homology sphere $\Sigma^+(t)$ and the collection of lens spaces
$L(|p'_s(t)|,q'_s(t))$ depend on the cosmological time $t$. Note
that among the lens spaces forming the boundaries of different
cobordisms $M^+_D(t)$ there exist mutually homeomorphic, namely
$L(a_1^{(l)},b_1^{(l)})$, since $a_1^{(l)}=a_1$ for $\forall
l\in\mathbb N$. By means of successive pairwise gluing together
these lens spaces, it is possible to form a cobordism $M_{\rm
total}$ which connects the initial state of universe, with a
spatial section $\Sigma^+(-4)$, to the final one with a spatial
section $\Sigma^+(0)$. The cobordism $M_{\rm total}$ will include
all intermediate stages with the following sequence of spatial
sections \be \label{4.14}
\Sigma^+(-4)\rightarrow\Sigma^+(-3)\rightarrow\Sigma^+(-2)
\rightarrow\Sigma^+(-1)\rightarrow\Sigma^+(0). \ee This is
accompanied by creation and annihilation of a certain set of
disjoint lens spaces of the type $L(|p'_s(t)|,q'_s(t))$ which have
no homeomorphic counterparts. This procedure is outlined in
\cite{EM}.

\subsection{Discussion: coupling constants of fundamental
interactions as cosmological circumstances} \label{s4.3}

In order to deepen our physical discussion, we give below the
calculations results for rational intersection matrices
$Q^{+IJ}(t)$ ($I,J=\overline{1,5+t}$), their eigenvalues, and
determinants, corresponding to graph cobordisms $M^+_D(t)$:

$$Q^{+IJ}(0)=\left(
\begin{array}{lllll}
{\bf 9.7\times 10^{\text{-1}}} & 3.1\times 10^{\text{-2}}
& 0 & 0 & 0 \\
3.1\times 10^{\text{-2}} & {\bf 7.2\times 10^{\text{-3}}}
& 1.4\times 10^{\text{-8}} & 0 & 0 \\
0 & 1.4\times 10^{\text{-8}} & {\bf 1.8\times 10^{\text{-12}}}
& 1.9\times 10^{\text{-29}} & 0\\
0 & 0 & 1.9\times 10^{\text{-29}} & {\bf 3.7\times 10^{\text{-44}}}
& 3.1\times 10^{\text{-89}} \\
0 & 0 & 0 & 3.1\times 10^{\text{-89}} & {\bf 2.7\times
10^{\text{-134}}}
\end{array}
\right),$$
$$\lambda^+_I(0)= \left\{9.7\times 10^{\text{-1}},6.2\times
10^{\text{-3}},1.7\times
10^{\text{-12}},3.7\times 10^{\text{-44}},6.4\times
10^{\text{-139}}\right\},$$
$$\det Q^{+IJ}(0)= 2.4\times10^{-196};$$
$$Q^{+IJ}(-1)=\left(
\begin{array}{llll}
8.3\times 10^{\text{-2}} & 1.1\times 10^{\text{-4}} & 0 & 0 \\
1.1\times 10^{\text{-4}} & 9.6\times 10^{\text{-6}} &
1.2\times 10^{\text{-14}} & 0 \\
0 & 1.2\times 10^{\text{-14}} & 2.5\times 10^{\text{-21}}
& 2.0\times 10^{\text{-44}} \\
0 & 0 & 2.0\times 10^{\text{-44}} & 1.6\times 10^{\text{-67}}
\end{array}
\right),$$
$$\lambda^+_I(-1)=\left\{8.3\times 10^{\text{-2}},9.5\times
10^{\text{-6}},2.5\times
10^{\text{-21}},9.7\times 10^{\text{-72}}\right\},$$
$$\det Q^{+IJ}(-1)=1.9\times10^{-98};$$
$$Q^{+IJ}(-2)=\left(
\begin{array}{lll}
3.0\times 10^{\text{-3}} & 1.5\times 10^{\text{-7}} & 0 \\
1.5\times 10^{\text{-7}} & 6.6\times 10^{\text{-10}} & 5.1\times
10^{\text{-22}} \\
0 & 5.1\times 10^{\text{-22}} & 4.0\times 10^{\text{-34}}
\end{array}
\right),$$
$$\lambda^+_I(-2)=\left\{3.0\times 10^{\text{-3}},6.5\times
10^{\text{-10}},7.9\times
10^{\text{-37}}\right\},$$
$$\det Q^{+IJ}(-2)= 1.5\times10^{-48};$$
$$Q^{+IJ}(-3)=\left(
\begin{array}{ll}
2.2\times 10^{\text{-6}} & 2.1\times 10^{\text{-12}} \\
2.1\times 10^{\text{-12}} & 2.2\times 10^{\text{-17}}
\end{array}
\right),$$
$$\lambda^+_I(-3)=\left\{2.2\times 10^{\text{-6}},2.0\times
10^{\text{-17}}\right\},$$
$$\det Q^{+IJ}(-3)= 4.4\times10^{-23};$$
$$Q^{+IJ}(-4)=\left(4.5\times 10^{\text{-9}}\right).$$
(In Appendix we shall give the rational intersection matrices
$Q^{-IJ}(t)$ corresponding to the cobordisms $M^-_D(t)$ obtained
from the splice diagrams in figure 9 by an application of the
$k^-$-operation while minimizing the parameter $k^-$ at each
step.)

Note that all elements $Q^{+IJ}(t)$ in these matrices are
rational; they are given here up to two significant digits. The
inverse matrices $Q^+_{IJ}(t)$ are integer. The inversion of the
rational intersection matrices with the help of the MAPLE program
is an excellent test of the correctness of their calculation
according to algorithms described in \cite{SavArt, EisNeum}, since
any error leads to non-integer elements in resulting matrices
$Q^+_{IJ}(t)$. Recalling the interpretation of rational
intersection matrices $\lambda Q^{+IJ}(t)$ as the coupling
constants of ``electric'' fluxes proposed in subsection \ref{s3.2}
(see {\it Observation 3.3}) we observe that the diagonal elements
of $5\times5$ matrix $Q^{+IJ}(0)$ (see boldface numbers) reproduce
rather exactly the hierarchy of DLEC constants for the well known
five fundamental interactions (see the fifth column in the table
2). The eigenvalues of this matrix reveal the same hierarchy. This
enables us to consider the interactions between ``electric''
fluxes $\Phi^{\rm(el)}_I(\Bar{m},\underline{l})$ defined in
(\ref{3.23}), see also their interpretation after (\ref{3.31}), as
``pre-images'' of the real fundamental interactions (or elementary
pre-interactions \cite{EM}). Then in accordance with table 2 we
shall relate the matrix elements $Q^{+II}(0)$ to strong (for
$I=1$), electromagnetic ($I=2$), weak ($I=3$), gravitational
($I=4$), and cosmological ($I=5$) pre-interactions. In this sense
the ``electric'' fluxes in the BF-model acquire the status of
quantized pre-fields bearing these names, {\it e.g.},
$\Phi^{\rm(strong)}
(\Bar{m},\underline{l}):=\Phi^{\rm(el)}_1(\Bar{m},\underline{l})$,
$\Phi^{\rm(electromagnetic)}
(\Bar{m},\underline{l}):=\Phi^{\rm(el)}_2(\Bar{m},\underline{l})$,
and so on.

It is natural to suppose that diagonal elements of the other
rational intersection matrices $Q^{+IJ}(t)$, $t\in
\overline{-4,-1}$ have hierarchy of the vacuum-level coupling
constants of the fundamental interactions (pre-interactions)
acting at earlier phases of cosmological evolution (which
correspond to the spacetime manifolds modeled by cobordisms
$M^+_D(t)$). Thus our model includes a certain unification scheme
of pre-interactions. So the intersection matrix $Q^{+IJ}(-1)$ has
the rank 4 and hence it describes the stage of universe with four
fundamental pre-interactions. This stage can be associated with
higher density of vacuum energy under which the topological
structure of the universe is reconstructed. But it would be too
speculative to directly connect this ``unification'' with the
electroweak unification theory, since in our model five
pre-interactions (between``electric'' fluxes) are replaced by
rather different (at least in the sense of hierarchy) four
pre-interactions.

With the same reservations one can relate the $3\times3$  matrix
$Q^{+IJ}(-2)$ to grand unified theories (GUT) in ordinary gauge
terms. The next $2\times2$  matrix $Q^{+IJ}(-3)$ may be associated
with a supersymmetric unification including the gravitation, since
out of five low-energy (for $t=0$) pre-interactions there survive
only two of them which correspond to gravitational and
cosmological pre-interactions. In this case the cobordism
$M^+_D(-3)$ should pertain to the Planck scales. Then the
$1\times1$ matrix (one rational number) $Q^{+IJ}(-3)$ might belong
to the sub-Planckian level where only one pre-interaction
(pre-image of the cosmological one) remains. It is obvious that in
order these interrelations might have some sense, we should first
introduce metric structures on the graph cobordisms $M^+_D(t)$,
then constructing over them field theories with local degrees of
freedoms. But we do not pose such a vast problem in this paper.

Now, let us see why manifestations of the presence of the
exceptional orbits in $\mathbb{Z}$-homology spheres (which are
spatial sections of our cosmological model), could be unobservable
by astronomical means. The idea is essentially the same as in the
inflation theory: to show that the linear scales of the
present-epoch $\mathbb{Z}$-homology sphere $\Sigma^+(0)$, are by
many orders of magnitude larger than the characteristic size of
the observable part of the universe ($L_0\sim 10^{28}$cm). To
evaluate the universe scales we take the following presumption.
Let the four-dimensional volume of universe $M^+(t)$ be
proportional to $\det Q^+_{IJ}(t)$ while the ``minimal volume'' in
this universe be proportional to $\det Q^{+IJ}(t)$. Then the
universe volume $V^+(t)$ expressed in terms of the ``minimal
volume'' should be $\det Q^+_{IJ}(t)/\det Q^{+IJ}(t)=\left(\det
Q^{+IJ}(t) \right)^{-2}$ which yields the expression for the
linear size of the universe as
$$L^+(t)\simeq\sqrt[4]{V^+(t)}=1/\sqrt{\det Q^{+IJ}(t)}.$$
Numerical estimates give the following results: \be \label{4.15}
\left. \begin{array}{l} L^+(-4)\sim 1.5 \times 10^4\\ L^+(-3)\sim
1.5 \times 10^{11}\\ L^+(-2)\sim 8.0 \times 10^{23}\\ L^+(-1)\sim
7.2 \times 10^{48}\\ \phantom{-}L^+(0)\sim 6.4 \times 10^{97}
\end{array} \right\}. \ee

As it was mentioned above, the state of universe corresponding to
$t=-3$ may be associated with a supersymmetric unification which
includes the gravitational pre-interaction. If linear scales of
the universe in this state might be considered as Planckian ones
($L_{Pl}\simeq 1.6\times 10^{-33}$cm), then the hierarchy
(\ref{4.15}) would be expressed in centimeters: \be \label{4.16}
\left. \begin{array}{l} L^+(-4)\sim 1.6 \times 10^{-40}~{\rm cm}\\
L^+(-3)\sim 1.6 \times 10^{-33}~{\rm cm}~~ {\rm (normalization)}\\
L^+(-2)\sim 8.6 \times 10^{-21}~{\rm cm}\\ L^+(-1)\sim 7.7 \times
10^{4}~~~{\rm cm}\\ \phantom{-}L^+(0)\sim 6.8 \times 10^{53}~~{\rm
cm} \end{array} \right\}. \ee These estimates give a plausible
picture of expansion of the universe in the course of cosmological
evolution. Four periods of moderate inflation take place,
$$1.6 \times 10^{-40}\rightarrow1.6 \times 10^{-33}\rightarrow8.6
\times 10^{-21}
\rightarrow 7.7 \times 10^{4}\rightarrow6.8 \times 10^{53},$$
which correspond to the sequence of topology changes (\ref{4.14}).
The size of the universe after the last ``inflation'' ($\sim 6.8
\times 10^{53}~~{\rm cm}$) occurs to be 25 orders of magnitude
greater then the size of its part which is observed now by means
of the most sophisticated astronomical devices. These evaluations
coincide with those obtained in $T_0$-discrete cosmological model
\cite{EMH04} except for the last one.

Consequently, all that we astronomically
observe is a three-dimensional almost
flat disk about $10^{28}$ cm in diameter cut out of the $\mathbb
Z$-homology sphere whose characteristic size amounts $6.8 \times
10^{53}$ cm. But while astronomical observations then have nothing
to do with spacetime topology, the local experiments providing
information about the hierarchy of fundamental interactions (in
contrast to ordinary inflation models) tell in our model
sufficiently much about non-trivial topological structure of the
spacetime. BF systems on graph cobordisms hint that the hierarchy
of physical interactions originates at the global level (the
utmost topological generalization of the Mach principle), so that
the background vacuum (excitations-free) coupling constants
naturally occur to coincide with basic topological invariants
(intersection matrices) of the spacetime manifold. It is clear
that the pre-interactions between ``electric'' fluxes in the
framework of Abelian BF-model considered in this paper, cannot
comprehensively express specific characteristics of the real
fundamental interactions. However our model (in spite of exotic
structure of the spacetime manifold or even due to these exotica)
heuristically circumscribes certain properties of Nature.

\section*{Acknowledgments}

We are grateful to Nikolai Saveliev for fruitful discussions in
our daily meetings during his visit to the University of
Guadalajara organized in the framework of our Proyecto de Posgrado
en Ciencias en F{\'\i}sica. We thank Gustavo L\'opez Vel\'azquez
for his interest and stimulating questions.

\section*{Appendix}

In this appendix we present the rational intersection matrices
$Q^-(t)$ corresponding to the cobordisms $M^-_D(t)$ obtained from
the splice diagrams in figure 9 applying the $k^-$-operation:

$$Q^-(0)=\left(
\begin{array}{lllll}
{\bf 1} & 3.6\times 10^{\text{-2}} & 0 & 0 & 0 \\
3.6\times 10^{\text{-2}} & {\bf 3.6\times 10^{\text{-2}}}
& 9.4\times 10^{\text{-8}} & 0 & 0 \\
0 & 9.4\times 10^{\text{-8}} & {\bf 1.9\times 10^{\text{-7}}}
& 2.1\times 10^{\text{-24}} & 0\\
0 & 0 & 2.1\times 10^{\text{-24}} & {\bf 1.5\times
10^{\text{-23}}} & 1.3\times 10^{\text{-68}} \\
0 & 0 & 0 & 1.3\times 10^{\text{-68}} & {\bf 1.1\times
10^{\text{-113}}}
\end{array}
\right),$$
$$\lambda^-_I(0)=\left\{1.,3.4\times 10^{\text{-2}},1.9\times
10^{\text{-7}},1.5\times 10^{\text{-23}},8.\times
10^{\text{-139}}\right\};$$
$$Q^-(-1)=\left(
\begin{array}{llll}
8.3\times 10^{\text{-2}} & 1.1\times 10^{\text{-4}} & 0 & 0 \\
1.1\times 10^{\text{-4}} & 4.3\times 10^{\text{-4}} & 5.5\times
10^{\text{-13}} & 0 \\
0 & 5.5\times 10^{\text{-13}} & 3.8\times 10^{\text{-12}}
& 3.1\times 10^{\text{-35}} \\
0 & 0 & 3.1\times 10^{\text{-35}} & 2.5\times 10^{\text{-58}}
\end{array}
\right),$$
$$\lambda^-_I(-1)=\left\{8.3\times 10^{\text{-2}},4.3\times
10^{\text{-4}},3.8\times 10^{\text{-12}},9.8\times
10^{\text{-72}}\right\};$$
$$Q^-(-2)=\left(
\begin{array}{lll}
3.0\times 10^{\text{-3}} & 1.5\times 10^{\text{-7}} & 0 \\
1.5\times 10^{\text{-7}} & 2.0\times 10^{\text{-6}}
& 1.5\times 10^{\text{-18}} \\
0 & 1.5\times 10^{\text{-18}} & 1.2\times 10^{\text{-30}}
\end{array}
\right),$$
$$\lambda^-_I(-2)=\left\{3.0\times 10^{\text{-3}},2.0\times
10^{\text{-6}},8.0\times 10^{\text{-37}}\right\};$$
$$Q^-(-3)=\left(
\begin{array}{ll}
2.2\times 10^{\text{-6}} & 2.1\times 10^{\text{-12}} \\
2.1\times 10^{\text{-12}} & 2.2\times 10^{\text{-17}}
\end{array}
\right),$$
$$\lambda^-_I(-3)=\left\{2.2\times 10^{\text{-6}},2.0\times
10^{\text{-17}}\right\};$$
$$Q^-(-4)=\left(4.5\times 10^{\text{-9}}\right).$$

It is easy to note that both diagonal elements of the matrix
$Q^-(0)$ and its eigenvalues  show another hierarchy then that of
the DLEC constants of the fundamental interactions (compare the
boldface numbers in this matrix with the ``experimental'' column
of the table 2 and with the corresponding characteristics of the
matrix $Q^+(0)$). So the matrices $Q^-(t)$ may be related to some
other universe, not with our one.

\end{document}